\documentclass[10pt, conference, compsocconf]{IEEEtran}
\newcommand{\ignore}[1]{}
\usepackage{epsfig}

\usepackage[pass]{geometry}
\usepackage{fancyhdr}
\usepackage[normalem]{ulem}
\usepackage[hyphens]{url}
\usepackage{hyperref}
\usepackage{color}
\usepackage{soul}

\usepackage{booktabs}
\usepackage{tabularx}
\usepackage{multirow}
\usepackage[sort,nocompress]{cite}
\usepackage[final]{microtype}
\usepackage{algorithm}
\usepackage{algpseudocode}
\usepackage{enumitem}

\usepackage{subcaption}
\newcommand{\Rebuttal}[1]{\textcolor{black}{#1}}
\newcommand{\fixme}[1]{\textcolor{black}{#1}}
\newcommand{\HPCA}[1]{\textcolor{black}{#1}}
\newcommand{\Camera}[1]{\textcolor{black}{#1}}
\usepackage{blindtext}

\fancypagestyle{firstpage}{
  \fancyhf{}
\setlength{\headheight}{50pt}

  \fancyhead[C]{\normalsize{Preprint for IEEE International Symposium on High-Performance Computer Architecture (HPCA 2020).
}}
}

\begin{document}
\title{The Architectural Implications of Facebook's \\ DNN-based Personalized Recommendation}

\author{\normalsize Udit Gupta\footnotemark[1], Carole-Jean Wu,  
Xiaodong Wang, Maxim Naumov, Brandon Reagen \\ \\
\normalsize David Brooks\footnotemark[1], Bradford Cottel, Kim Hazelwood, Mark Hempstead, Bill Jia, Hsien-Hsin S. Lee, Andrey Malevich,  \\ 
\normalsize Dheevatsa Mudigere, Mikhail Smelyanskiy, Liang Xiong, Xuan Zhang \\ \\
\normalsize Facebook Inc.\\
\normalsize \{carolejeanwu, xdwang\}@fb.com
}

\maketitle
\addtocounter{footnote}{+1}

\footnotetext{Harvard University, work done while at Facebook.}
\thispagestyle{firstpage}


\begin{abstract}

The widespread application of deep learning has changed the landscape of computation in data centers.
In particular, personalized recommendation for content ranking
is now largely accomplished using deep neural networks.
However, despite their importance and the amount of compute cycles they
consume, relatively little research attention has been devoted to recommendation systems.
To facilitate research and advance the understanding of these workloads, 
this paper presents a set of real-world, production-scale DNNs for personalized recommendation 
coupled with relevant performance metrics for evaluation.
In addition to releasing a set of open-source workloads, 
we conduct in-depth analysis that underpins future system design and optimization for at-scale recommendation: 
Inference latency varies by 60\% across three Intel server generations, 
batching and co-location of inference jobs can drastically improve latency-bounded throughput, 
and diversity across recommendation models leads to different optimization strategies.

\end{abstract}

\section{Introduction}

Deep learning has become a cornerstone in many
production-scale data center services.
As web-based applications continue to expand globally, 
so does the amount of compute and storage resources
devoted to deep learning training and 
inference~\cite{chung2018serving, tpu, hazelwood2018applied}.
Personalized recommendation is an important class of these services.
Deep learning based recommendation systems
are broadly used throughout industry to predict rankings for
news feed posts and entertainment content~\cite{youtube, netflix}. 
For instance, in 2018, McKinsey and Tech Emergence estimated that recommendation systems were
responsible for driving up to 35\% of Amazon's revenue~\cite{chui2018notes,microsoftPersonalizedRec,mckinsey}.

\begin{figure}[t!]
  \centering
  \includegraphics[width=\columnwidth]{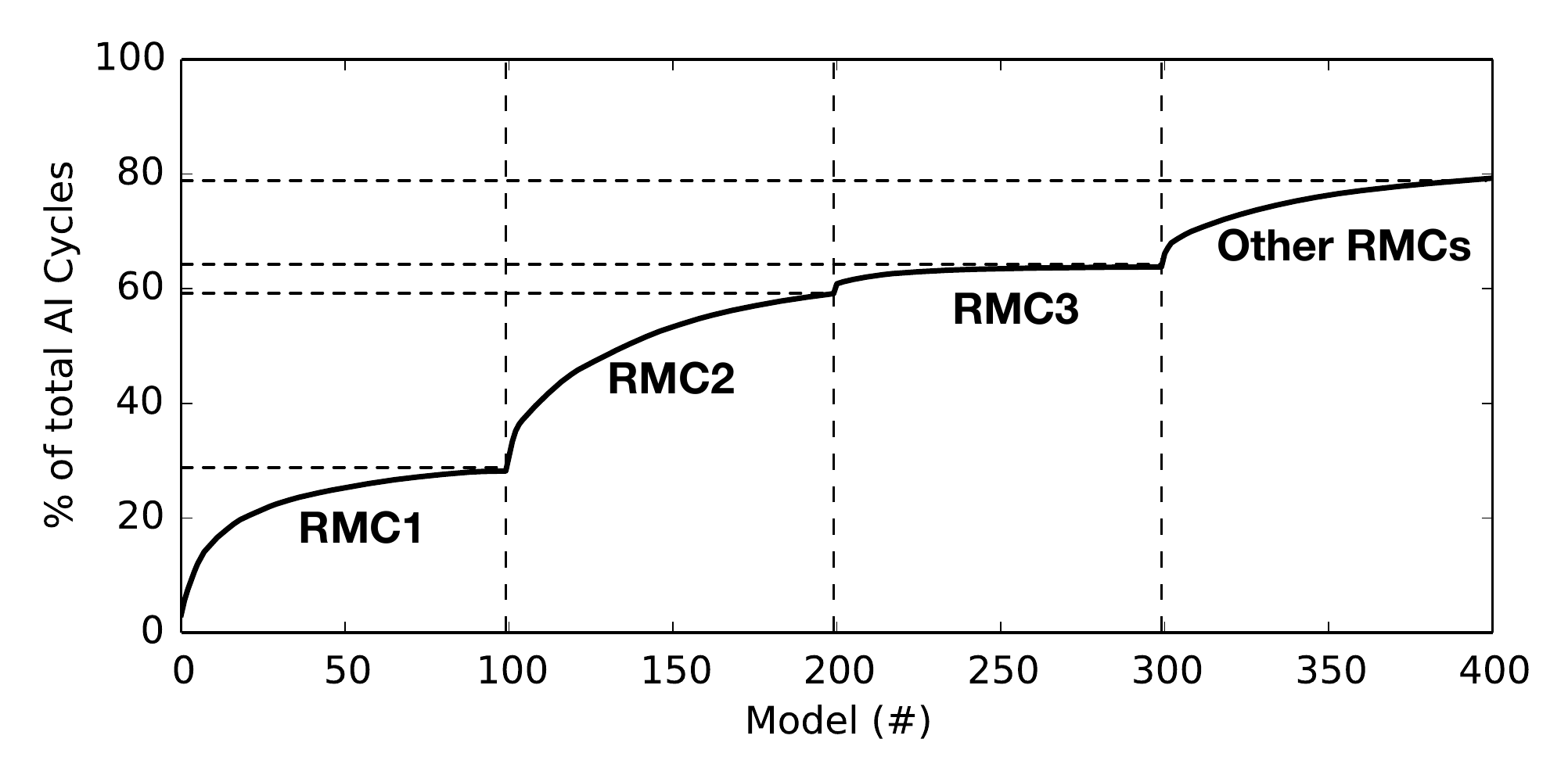}
  \vspace{-2em}
  \caption{
  \HPCA{RMC1, RMC2, and RMC3 (studied in this paper) represent three classes of recommendation models that consume 65\% of AI inference cycles in Facebook's production data center. 
  Recommendation models in general comprise over 79\% of AI inference cycles. 
  Remaining cycles are devoted to non-recommendation use cases (e.g., CNN, RNNs).}
  }
  \vspace{-1em}
  \label{fig:model_dist}

\end{figure}

\HPCA{Figure \ref{fig:model_dist} illustrates the fraction of AI inference cycles spent across recommendation models in a production data center. 
DNN-based personalized recommendation models comprise up to 79\% of AI inference cycles in a production-scale data center. 
While potentially hundreds of recommendation-models are used across the data center, we find that three \underline{r}ecommendation \underline{m}odel \underline{c}lasses, RMC1, RMC2, and RMC3, consume up to 65\% of AI inference cycles.
These three types of recommendation models follow distinct recommendation model architectures that result in different performance characteristics and hardware resource requirements, and thus are the focus of this paper.}

The systems and computer architecture community has made significant strides
in optimizing the performance, energy efficiency, and memory consumption of
DNNs.
Recent solutions span across the entire system stack including, efficient DNN architectures~\cite{resnet,howard2017mobilenets,chung2014empirical}, reduced precision datatypes~\cite{han2015deep,minerva,judd2016stripes,courbariaux2016binarized,gupta2015deep}, heavily parallelized training/inference~\cite{goyal2017accurate,you2018imagenet}, and hardware accelerators~\cite{tpu, minerva, eyeriss, eie, cambricon-x}.
These solutions primarily target 
convolutional (CNN)~\cite{resnet, yolo}
and recurrent (RNN)~\cite{ds2, ds3} neural networks.
However, \textit{these optimization techniques often cannot be applied
to recommendation models} as the models are intrinsically different,
introducing unique memory and computational challenges.


\begin{figure}[t!]
  \centering
  \includegraphics[width=\columnwidth]{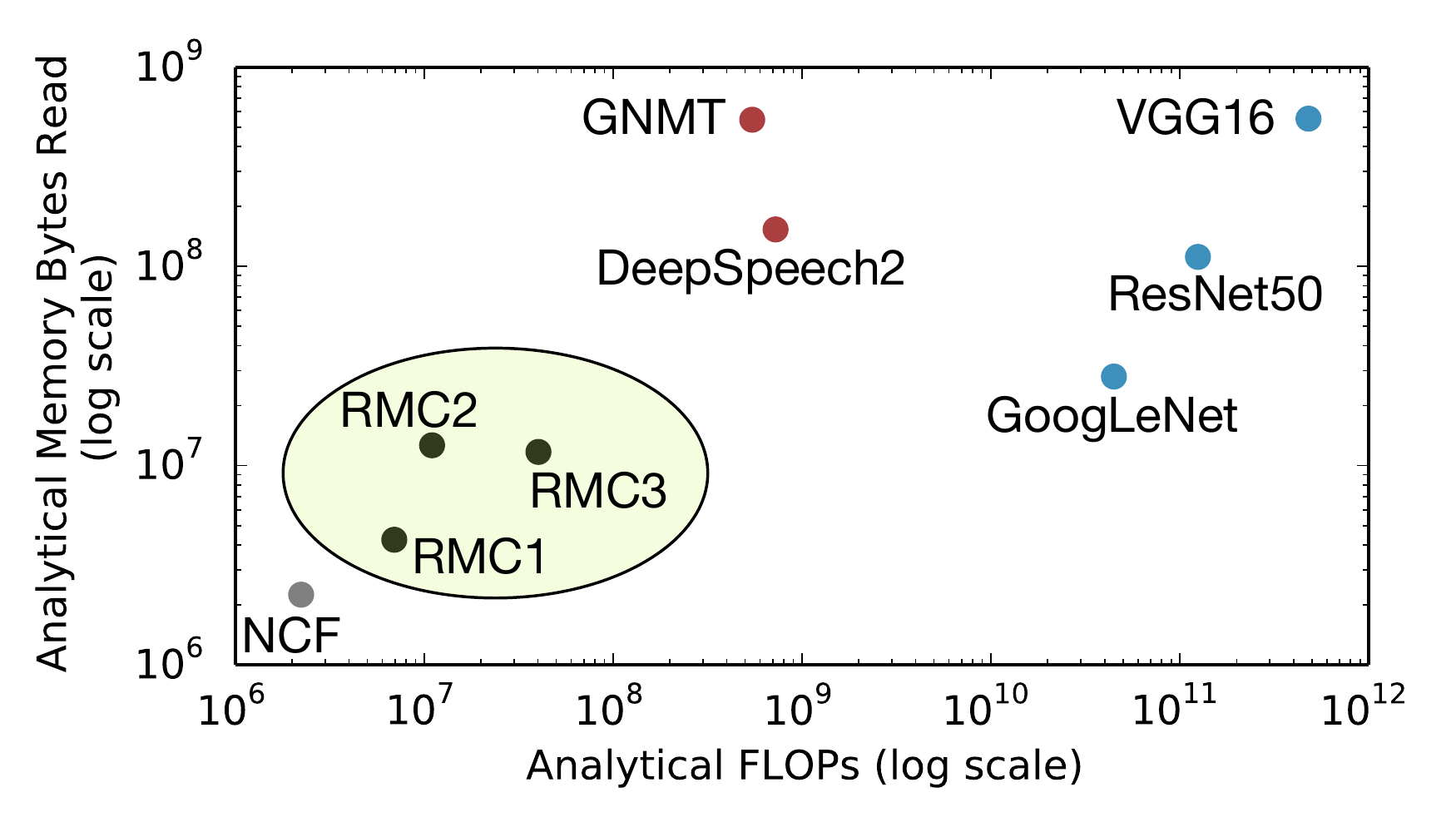}
  \vspace{-1em}
  \caption{  \HPCA{At-scale recommendation models (RMC1, RMC2, RMC3) have unique compute (FLOPs) and memory (bytes read) requirements compared to CNNs, RNNs, and open-source recommendation models like MLPerf-NCF~\cite{mlperf,mlperf-training}. }   }
  \vspace{-1.0em}
  \label{fig:analytical_networks}

\end{figure}

Finally, publicly available recommendation benchmarks are not representative of production systems.
Compared to available recommendation benchmarks, i.e. neural-collaborative filtering (MLPerf-NCF~\cite{mlperf,mlperf-training}), production-scale models differ in three important features: \fixme{application-level constraint} (use case), embedding tables (memory intensive), fully-connected layers (compute intensive).
\fixme{First, production recommendation use cases require processing requests with high throughput under strict latency constraints; to meet these application requirements, production systems exploit high degrees of data-level and task-level parallelism not considered in publicly available benchmarks.}
Second, production-scale recommendation models have \textit{orders of magnitude} more embeddings, resulting in larger storage requirements and more irregular memory accesses. 
Finally, MLPerf-NCF implements fewer and smaller fully-connected (FC) layers requiring fewer FLOPs (Figure~\ref{fig:analytical_networks}).
The insights and solutions derived using these smaller recommendation models may not be applicable to nor representative of production systems.

In this paper, we present a set of production-scale personalized recommendation models. 
First, we identify quantitative metrics to evaluate 
the performance of these recommendation workloads.
Next, we design a set of synthetic recommendation models
to conduct detailed performance analysis. 
Because inference in our data center is run across a variety of
CPUs~\cite{hazelwood2018applied}, 
we focus the design tradeoff studies on 
Intel Haswell, Broadwell, and Skylake servers. 
Finally, we study performance characteristics of running recommendation models in production-environments. 
The insights from this analysis can be used to motivate broader system and 
architecture optimizations for at-scale recommendation. 
For example, we can maximize latency-bounded throughput by exploiting server heterogeneity when scheduling inference requests. 

The in-depth description and characterization presented in this paper
of production-scale recommendation models provides 
the following insights for future system design:

\begin{itemize}[leftmargin=*]
  \item The current practice of using only latency for benchmarking inference performance is insufficient. 
  At the data center scale, the metric of latency-bounded throughput is more representative as it determines the number of items that can be ranked given service level agreement (SLA) requirements (Section~\ref{sec:SparseNNParams}). 

  \item Inference latency varies across several generations of Intel servers (Haswell, Broadwell, Skylake) that co-exist in data centers.
    With unit batch size, inference latency is optimized on high-frequency Broadwell machines.
    On the other hand, batched inference (throughput) is optimized with Skylake as batching increases the compute density of FC layers.
    Compute-intensive recommendation models are more readily accelerated with AVX-512 instructions in Skylake, as compared to AVX-2 in Haswell and Broadwell (Section~\ref{sec:single}).
    

  \item Co-locating multiple recommendation models on a single machine can improve throughput.
    However, this introduces a tradeoff between single model latency and aggregated system throughput. 
    We characterize this tradeoff and find that processors with inclusive L2/L3 cache hierarchies (i.e., Haswell, Broadwell) are particularly susceptible to latency degradation due to co-location.
    This introduces additional scheduling optimization opportunities in data centers  (Section~\ref{sec:colocation}).
    
      \item Across at-scale recommendation models and different server architectures, the fraction of time spent on compute intensive operations, like FC, varies from 30\% to 95\%.
      Thus, existing solutions for accelerating FC layers only \cite{tpu, minerva, eie, cambricon-x} will translate to limited inference latency improvement for end-to-end recommendation.
      This is especially true of recommendation models dominated by embedding tables (Section~\ref{sec:single}).
\end{itemize}

\begin{figure}[t!]
  \centering
  \includegraphics[width=0.95\columnwidth]{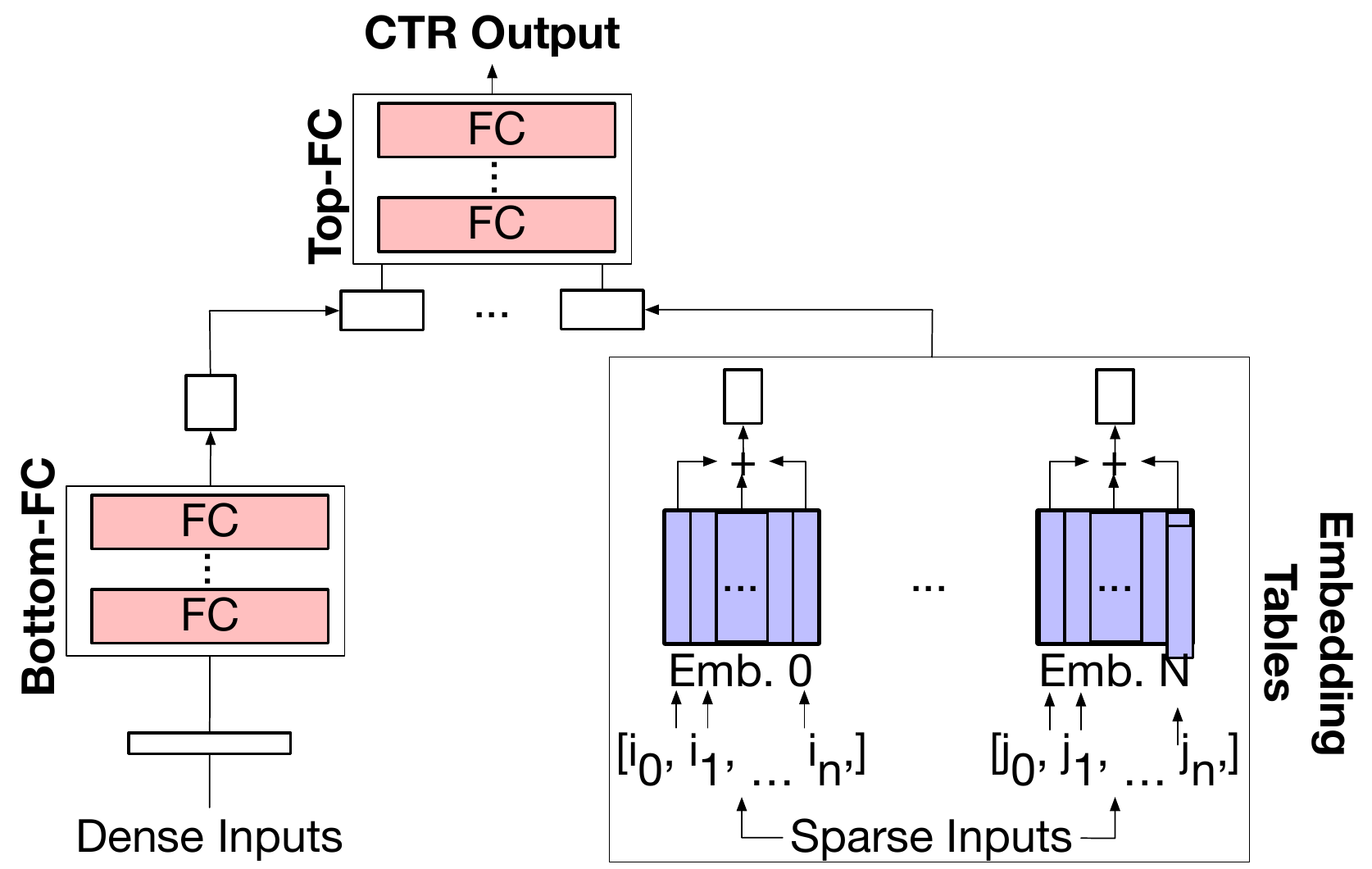}
  \caption{
  Simplified model-architecture of recommendation models. 
  Inputs to the model are a collection of dense and sparse features. 
  Sparse features, unique to recommendation models, are transformed to a
dense representation using embedding tables (blue). 
The number/size of embedding tables, number
of sparse feature (ID) lookups per table, depth/width of Bottom-FC and Top-FC layers varies based on the use-case.}
  \label{fig:sparsenn}
  \vspace{-1em}
\end{figure}

\textbf{Open-source:} To facilitate future work on at-scale recommendation systems for the systems and computer architecture community, \HPCA{Facebook} has open-sourced a suite of synthetic models, representative of production use cases\footnote{\href{https://ai.facebook.com/blog/dlrm-an-advanced-open-source-deep-learning-recommendation-model/}{https://ai.facebook.com/blog/dlrm-an-advanced-open-source-deep-learning-recommendation-model/}}. 
Together with the detailed performance analysis performed in this paper, the open-source implementations can be used to further understand the compute requirements, storage capacity, and memory access patterns, enabling optimization and innovation for at-scale recommendation systems.



\section{Background}

This section provides an overview of the personalized recommendation task and the architecture of at-scale recommendation models.
We also compare recommendation models to other DNNs, in terms of their compute density, storage capacity, and memory access patterns. 

\subsection{Recommendation Task}
Personalized recommendation is the task of recommending new content  to users based on their preferences ~\cite{youtube, netflix}.
Estimates show that up to 75\% of movies watched on Netflix and 60\% of videos consumed on YouTube are based on suggestions from their recommendation systems~\cite{chui2018notes,microsoftPersonalizedRec,mckinsey}.

Central to these services is the ability to accurately, and efficiently rank content based on users' preferences and previous interactions (e.g., clicks on social media posts, ratings, purchases).
Building highly accurate personalized recommendation systems poses unique challenges as user preferences and past interactions are represented as both \textit{dense} and \textit{sparse} features \cite{eisenman2018bandana,naumov2019dimensionality}.

\begin{figure}[t!]
  \centering
  \includegraphics[width=\columnwidth]{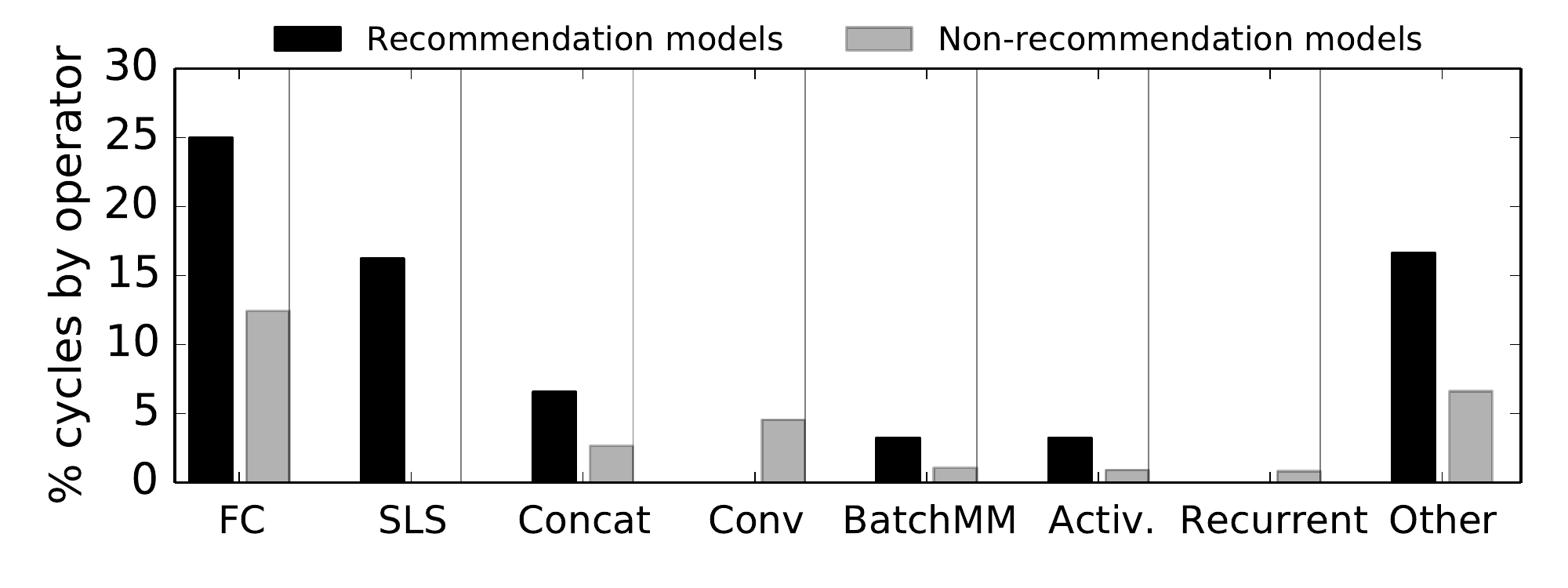}
  \vspace{-2em}
  \caption{\HPCA{Breakdown of data center-wide cycles by operators in recommendation and non-recommendation (e.g., CNN, RNNs) models.}}
  \vspace{-1.25em}
  \label{fig:op_dist}
\end{figure}


For instance, in the case of ranking videos (e.g., Netflix, YouTube), there may be tens of thousands of videos that have been seen by millions of viewers. 
However, individual users interact with only a handful of videos.
This means interactions between users and videos are sparse.
\Rebuttal{Sparse features typically represent categorical inputs. For example, for ranking videos, a categorical input may represent the type of device or users' preferences for a genre of content~\cite{wideanddeep}. 
Dense features (e.g., user age) represent continuous inputs.
Categorical inputs are encoded as multi-hot vectors where a $1$ represents positive interactions.
Given the potentially large domain (millions) and small number of interactions (sparse), multi-hot vectors must first be transformed into real-valued dense vectors using embedding table operations.}
Sparse features not only make training more challenging but also require intrinsically different operations (e.g., embedding tables) which impose unique compute, storage capacity, and memory access pattern challenges.


\subsection{Recommendation Models} \label{sec:recModels}
Figure \ref{fig:sparsenn} shows a simplified architecture of state-of-the-art DNNs for personalized recommendation models. 
(More advanced examples can be in found~\cite{instagramSparseNN,alibabaRec}.)
The model comprises a variety of operations such as FC layers, embedding tables (which transform sparse inputs to dense representations), Concat, and non-linearities, such as ReLU.
At a high-level, dense and sparse input features are separately transformed using FC layers and embedding tables respectively. 
The outputs of these transformations are then combined and processed by a final set of FC layers.
\HPCA{Figure \ref{fig:op_dist} illustrates the cycles breakdown of these operators across Facebook's data centers. 
Given their unique architecture, the cycle breakdown of recommendation models follows a distinct distribution compared to non-recommendation models (e.g., CNNs, RNNs). 
In particular, FC, SLS and Concat comprise over 45\% of recommendation cycles.
Note that, SLS (embedding table operations in Caffe2) alone comprise nearly 15\% of AI inference cycles across Facebook's data centers --- 4$\times$ and 20$\times$ more than CNNs and RNNs.}

\textbf{Execution Flow:} The inputs, for a single user and single post, to recommendation models are a set of dense and sparse features. 
The output is the predicted click-through-rate (CTR) of the user and post.
Dense features are first processed by a series of FC layers, shown as the Bottom-FCs in Figure \ref{fig:sparsenn}.
Recall that sparse features represent categorical inputs that can be encoded as multi-hot vectors.
As the number of categories (i.e., size of multi-hot vectors) is large, each vector is encoded as a list of non-contiguous, sparse IDs. 

\begin{figure}[t!]
  \centering
  \includegraphics[width=\columnwidth]{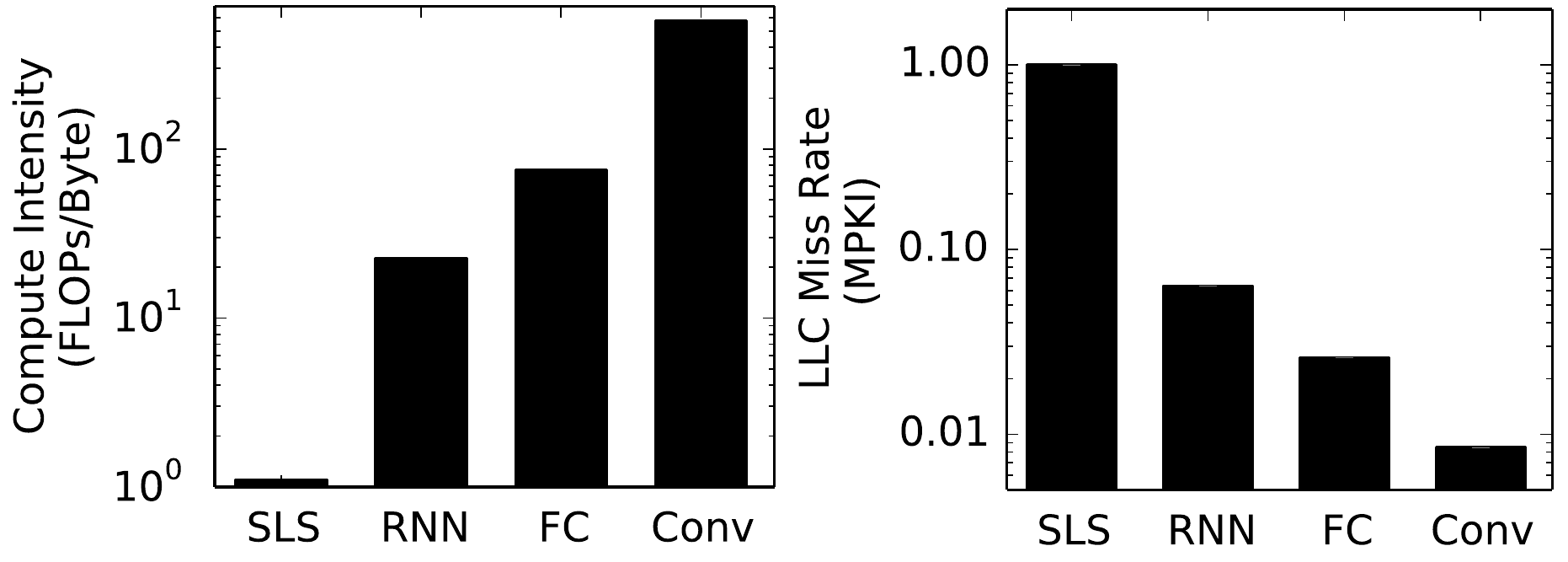}
  \vspace{-1em}
  \caption{Compared to FC, CNN, and RNN layers, embedding table operations (SparseLengthsSum, SLS, in Caffe2), seen in recommendation systems, exhibit low compute density (left) and high LLC cache miss rate (right).}
  \vspace{-1.5em}
  \label{fig:embtable}
\end{figure}

For a single user-item input pair, multiple vectors of such sparse IDs must first be made dense.
While the sparse to dense transformation can be accomplished using FC layers, the compute demands of doing so would be significant.
Instead, we use embedding tables.
Each vector is paired with an embedding table, as shown in Figure \ref{fig:sparsenn}, and each sparse ID is used to look-up a unique row in the embedding table. \Rebuttal{(Pseudo-code in Algorithm \ref{alg:sls}).}
The rows of the embedding are then combined into a single vector, typically with a dimension of 32 or 64, using element-wise operations.

Finally, embedding vectors and the output of the Bottom-FC layers are concatenated, and processed by the Top-FC layers shown in Figure \ref{fig:sparsenn}.
The output is a single value representing the predicted CTR. 
User-post pairs with the highest predicted CTR will be prioritized.

\textbf{Processing multiple posts:} 
At the data center scale, recommendations for many users and posts must be ranked simultaneously. 
Thus, it is important to note that the vectors of sparse IDs shown in Figure \ref{fig:sparsenn} correspond to inputs for a single user and single post.
To compute the CTR of many user-post pairs at once, requests are batched to improve overall throughput.
\begin{algorithm}[t]
\small
\caption{\HPCA{SparseLengthsSum (SLS) pseudo-code}}\label{alg:sls}
\begin{algorithmic}[1]
\State $Emb \gets \textrm{Embedding Table: \textbf{R}($\sim$millions) x \textbf{C}($\sim$tens)}$
\State $Lengths \gets \textrm{Vector: \textbf{K}}$ \Comment{slices of IDs\, \, \, \, }
\State $IDs \gets \textrm{Vector: \textbf{M} ($\sim$tens)}$ \Comment{non-contiguous\,\,}
\State $Out \gets Vector: \textbf{K} \times \textbf{C} $ 
\\
\State $CurrentID = 0; OutID = 0$
\Procedure{SLS}{Emb, Lengths, IDs}
\For{L in Lengths} 
\For{ID in IDS[CurrentID: CurrentID+L]:} 
\State $Emb\_vector = Emb[ ID ]$
\For{i in range(C):}
\State $Out[OutID][i] += Emb\_vector[i]$

\EndFor
\EndFor
\State $OutID = OutID + 1; CurrentID = CurrentID + L$
\EndFor
\State $\textbf{return}\,\,\,\, Out$
\EndProcedure
\end{algorithmic}
\end{algorithm}

The depth and width of FC layers, number and size of embedding tables, number of sparse IDs per input, and typical batch-sizes depend on the use case of the recommendation model (see Section \ref{sec:SparseNNParams} for more details). 

\subsection{Embedding Tables}
A key distinguishing feature of DNNs for recommendation systems, compared to CNNs and RNNs, is the use of embedding tables.
As shown in Figure \ref{fig:sparsenn}, embedding tables are used to transform sparse input features to dense ones.
However, the embedding tables impose unique challenges to efficient execution in terms of their large \textit{storage capacity}, \textit{low compute density}, and \textit{irregular memory access} pattern.

\textbf{Large storage capacity} 
The size of a single embedding table seen in production-scale recommendation models varies from tens of MBs to several GBs.
Furthermore, the number of embedding tables varies from 4 to 40, depending on the particular use case of the recommendation model.
(See Section \ref{sec:SparseNNParams} for details).
In aggregate, embedding tables for a single recommendation model can consume up to tens of GB of memory.
Thus, systems running production-scale recommendation models require large, off-chip storage such as DRAM or dense non-volatile memory \cite{eisenman2018bandana}.


\textbf{Low compute intensity}
As shown in Figure \ref{fig:embtable}(left), SparseLengthsSum (SPS) has a significantly lower compute intensity, e.g. operational  intensity, (0.25 FLOPs/Byte) compared to RNN (5.5 FLOPs/Byte), FC (18 FLOPs/Byte), and CNN (141 FLOPs/Byte) layers. 
As shown in Figure \ref{fig:embtable}, compared to typical FC, RNN, and CNN layers, embedding tables exhibit low compute density.
(The RNN layer considered is typically found in recurrent NLP models while the FC and CNN layers are ones found in ResNet50~\cite{resnet}). 
Recall that the embedding table operation (implemented as the \textit{SparseLengthsSum} operator in Caffe2\cite{caffe2}) entails reading a small subset of rows in the embedding table. 
The rows, indexed based on input sparse IDs, are then summed.
While the entire embedding table is not read for a given input, the accesses follow a highly irregular memory access pattern \Rebuttal{(Pseudo-code in Algorithm \ref{alg:sls})}.

\begin{figure}[t!]
  \centering
  \includegraphics[width=0.95\columnwidth]{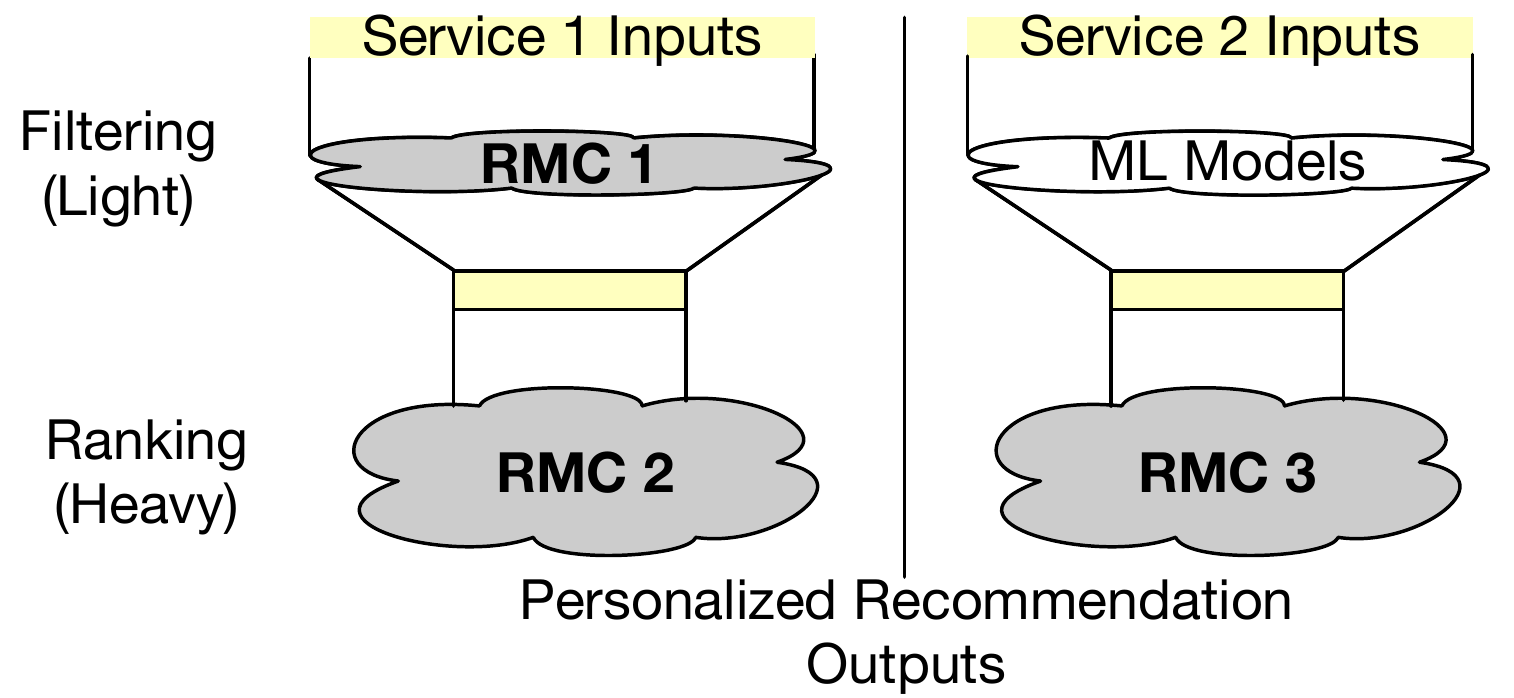}
  \caption{Contents is ranked
in two hierarchical steps: filtering and ranking. Filtering reduces the number of total items to a smaller subset using lightweight machine learning techniques or smaller
DNN-based models (RMC1). 
Ranking performs finer grained recommendation using larger models (e.g., RMC2, RMC3).}
  \vspace{-1em}
  \label{fig:flow}

\end{figure}

\textbf{Irregular memory accesses}
On an Intel Broadwell server present in production data centers, this results in a high LLC cache miss rate.
For instance, Figure \ref{fig:embtable}(right) shows that a typical SparseLengthsSum operator in production-scale recommendation models has an LLC cache miss rate of 8 MPKI \cite{eisenman2018bandana}, compared to 0.5 MPKI, 0.2 MPKI, and 0.06 MPKI in an RNN, FC, and CNN layers. 
Previous work~\cite{eisenman2018bandana} demonstrates that embedding table lookups exhibit low reuse; high miss-rates are a result of compulsory misses not capacity misses. 
The observed cache miss rates can be exacerbated by additional interference from OS-related activities, processor-dependent TLB miss handling, as well as prefetching pollution~\cite{Wu:ISPASS11,Ebrahimi:MICRO09,Wu:MICRO11}.
Furthermore, the element-wise sum is a low-compute intensity operation. Due to their highly irregular memory access pattern and low-compute density, efficient embedding table operations requires unique solutions, compared to  approaches applied to FC and CNN layers.

\begin{table*}[t!]
\begin{center}
\begin{tabular}{|c|l||c|c|l|l|c|c|}
\hline
\multirow{2}{*}{\textbf{Model}} & \multirow{2}{*}{\textbf{Description}} & \multicolumn{2}{|c|}{\textbf{FC}} & \multicolumn{4}{|c|}{\textbf{Embedding Tables}}  \\

        &  & Bottom & Top & Number & Input Dim. & Output Dim. & Lookups \\ \hline \hline                                                                                                                                                            
                                                                                                                                                                                                                                                              
        \multirow{3}{*}{RMC1}                    & Small FC          & Layer1: 8$\times$  & Layer1: 4$\times$ & \multirow{3}{*}{1$\times$ to 3$\times$}  & \multirow{3}{*}{1$\times$ to 180$\times$}  & \multirow{3}{*}{1$\times$} & User: 4$\times$  \\       
                                                & Few Emb. Tables   & Layer2: 4$\times$  & Layer2: 2$\times$ &                                          &                                            &                            & Posts:Nx4$\times$ \\      
                                                & Small Emb. Tables & Layer3: 1$\times$  & Layer3: 1$\times$ &                                          &                                            &                            & \\ \hline                 
            \multirow{3}{*}{RMC2}                & Small FC          & Layer1: 8$\times$  & Layer1: 4$\times$ & \multirow{3}{*}{8$\times$ to 12$\times$} & \multirow{3}{*}{1$\times$ to 180$\times$}  & \multirow{3}{*}{1$\times$} & User:4$\times$  \\        
                                                & Many Emb. Tables  & Layer2: 4$\times$  & Layer2: 2$\times$ &                                          &                                            &                            & Posts:Nx4$\times$\\       
                                                & Small Emb. Tables & Layer3: 1$\times$  & Layer3: 1$\times$ &                                          &                                            &                            & \\ \hline                 
                \multirow{3}{*}{RMC3}            & Large FC          & Layer1: 80$\times$ & Layer1: 4$\times$ & \multirow{3}{*}{1$\times$ to 3$\times$}  & \multirow{3}{*}{10$\times$ to 180$\times$} & \multirow{3}{*}{1$\times$} & \multirow{3}{*}{1$\times$}\\                                                                                                                                                                                                                                                            
                                                & Few Emb. Tables   & Layer2: 8$\times$  & Layer2: 2$\times$ &                                          &                                            &                            & \\                        
                                                & Large Emb. Tables & Layer3: 4$\times$  & Layer3: 1$\times$ &                                          &                                            &                            & \\ \hline   

\end{tabular}                                                                                                                                                                                                                                                 
\end{center}                                    \caption{Model architecture parameters representative of production scale recommendation workloads for three example recommendation models used, highlighting their diversity in terms of embedding table and FC sizes. Each parameter (column) is normalized to the smallest instance across all three configurations. 
Bottom and Top FC sizes are normalized to layer 3 in RMC1. Number, input dimension, and output dimension of embedding tables are normalized to RMC1.
Number of lookups are normalized to RMC3.
}
\vspace{-1em}
\label{tab:RMParams}                                                                                                                                                                                                                                                                                                         
\end{table*}

\section{At-scale Personalization}
\label{sec:SparseNNParams}
This section describes model architectures for three classes of production-scale recommendation models: RMC1, RMC2, and RMC3.
\Rebuttal{This paper focuses on these three classes of recommendation models for two reasons. 
First, }
the models span different configurations in terms of the number and size of embedding tables, number of sparse IDs per embedding table, and size of FC layers.
These configurations determine compute density, storage requirements, and memory access patterns, which may lead to different system and micro-architecture optimizations.
\HPCA{Second, the models consume the 65\% of inference cycles in Facebook's data centers (Figure~\ref{fig:model_dist})}.


\subsection{Production Recommendation Pipeline}
As shown in Figure \ref{fig:flow}, personalized recommendation is accomplished by hierarchically ranking content.
Lets consider the example of recommending social media posts.
When the user interacts with the web-based social media platform, a request is made for relevant posts. 
At any given time there may be thousands of relevant posts.
Based on user preferences, the platform must recommend the top tens of posts.
This is accomplished in two steps, filtering and ranking~\cite{wideanddeep}.

First, the set of possible posts, thousands, is filtered down by orders of magnitude.
This is accomplished using lightweight machine learning techniques such as logistic regression. 
Compared to using heavier DNN-based solutions, using lightweight techniques trades off higher accuracy for lower run-time.
DNN-based recommendation models are used in the filtering step when higher accuracy is needed.
One such example is recommendation model 1 (RMC1). 

Next, the subset of posts is ranked and the top tens of posts are shown to the user.
This is accomplished using DNN-based recommendation models.
Compared to recommendation models used for filtering content, models for finer grained ranking are typically larger in terms of FC and embedding tables.
For instance, in the case of ranking social media posts, the heavyweight recommendation model (i.e., RMC3) is comprised of larger Bottom-FC layers.
This is a result of the service using more dense features.
The other class of heavyweight recommendation models (i.e., RMC2) is comprised of more embedding tables as it processes contents with more sparse features. 

\textbf{SLA requirements:} In both steps, lightweight filtering and heavyweight ranking, many posts must be considered per user query.
Each query must be processed within strict latency constraints set by service level agreements (SLA). 
\Rebuttal{For personalized recommendation, \textit{missing latency targets results in jobs being preemptively terminated, degrading recommendation result quality}. }
The SLA requirements can vary from tens to hundreds of milliseconds \cite{tpu, wideanddeep, park2018deep}.
Thus, when analyzing and optimizing recommendation systems in production data centers, it is important to consider not only single model latency but also throughput metrics under SLA.
In the data center, balancing throughput with strict latency requirements is accomplished by batching queries and co-locating inferences on the same machine (Section \ref{sec:single} and Section \ref{sec:colocation}).

\subsection{Diversity of Recommendation Models}

Table \ref{tab:RMParams} shows representative parameters for three classes of recommendation models: RMC1, RMC2, and RMC3. 
While all three types of models follow the general  architecture (Figure \ref{fig:sparsenn}), they are quite diverse in terms of number and size of embedding tables, embedding table lookups, and depth/width of FC layers.
To highlight these differences we normalize each feature to the smallest instance across all models.
Bottom and Top FC sizes are normalized to layer 3 in RMC1.
Number, input, and output dimensions of embedding tables are normalized to RMC1.
The number of lookups (\textit{sparse} IDs) per embedding table are normalized to RMC3. 
RMC1 is smaller in terms of FCs and embedding tables, RMC2 has many embedding tables (memory intensive), and RMC3 has larger FCs (compute intensive).
Note that the number of FC parameters in the Top-FC layer depends on not only the layer dimensions, but also the input size which scales with the number of embedding tables (Figure ~\ref{fig:sparsenn}) and potentially large.

\textbf{The number and size of embedding tables} across the three classes of recommendation models.
For instance, RMC2 can have up to an order of magnitude more embedding tables compared to RMC1 and RMC3. 
This is because RMC1 is a lightweight recommendation model used in the initial filtering step and RMC3 is used in applications with fewer sparse features. 
Furthermore, while the output dimension of embedding tables is the same across the recommendation models (between 24-40), RMC3 has the largest embedding tables in terms of the input dimensions.
In aggregate, assuming 32-bit floating point datatypes, the storage capacity of embedding tables varies between 100MB, 10GB, and 1GB for RMC1, RMC2, and RMC3. 
Thus, systems that run any of the three at-scale recommendation model types, require large, off-chip memory systems.

\textbf{Embedding table lookups}
Embedding tables in RMC1 and RMC2 have more lookups (i.e., more sparse IDs) per input compared to RMC3.
This is a result of RMC1 and RMC2 being used in services with many sparse features 
while RM3 is used in recommending social media posts, which has fewer sparse features.
Thus, RMC1 and RMC2 models perform more irregular memory accesses leading to higher cache miss rates on off-the-shelf Intel server architectures found in the data center.

\textbf{MLP layers}
Bottom-FC layers for RMC3 are generally much wider than those of RMC1 and RMC2.
This is a result of using more dense features in ranking social media posts (RMC3) compared to services powered by RMC1 and RMC2.
Thus, RMC3 is a more compute intensive model than RMC1 and RMC2.
Finally, it is important to note that width of FC layers is not necessarily a power of 2, or cache-line aligned.

\begin{table}[t!]
\begin{center}
\small
\begin{tabular}{|c||c|c|c|}
\hline
\textbf{Machines} & \textbf{Haswell} & \textbf{Broadwell} & \textbf{Skylake} \\ \hline
Frequency & 2.5GHz & 2.4GHz & 2.0GHz \\ \hline
Cores per socket & 12 & 14 & 20 \\ \hline
Sockets & 2 & 2 & 2 \\ \hline
SIMD & AVX-2 & AVX-2 & AVX-512 \\ \hline
L1 Cache Size & 32 KB & 32 KB & 32 KB \\ \hline
L2 Cache Size & 256 KB & 256 KB & 1MB \\ \hline
L3 Cache Size & 30 MB & 35 MB & 27.5MB \\ \hline 
L2/L3 Inclusive & \multirow{2}{*}{Inclusive} & \multirow{2}{*}{Inclusive} & \multirow{2}{*}{Exclusive} \\ 
or Exclusive & & & \\ \hline
DRAM Capacity & 256 GB & 256 GB & 256GB \\ \hline
DDR Type  & DDR3 & DDR4 & DDR4 \\ \hline
DDR Frequency  & 1600MHz & 2400MHz & 2666MHz \\ \hline
DDR Bandwidth & \multirow{2}{*}{51 GB/s} & \multirow{2}{*}{77 GB/s} & \multirow{2}{*}{85 GB/s} \\
per socket & & & \\ \hline
\end{tabular}
\end{center}
\vspace{-1em}
  \caption{ Description of machines present in data centers
and used to run recommendation models}
  \label{tab:machines}
  \vspace{-1em}
\end{table}
\section{Experimental Setup}
\textbf{Server Architectures:}
Generally, data centers are composed of a heterogeneous set of server architectures with differences in compute and storage capabilities. 
Services are mapped to racks of servers to match their compute and storage requirements.
For instance, ML inference in data centers is run on large dual-socket server-class Intel Haswell, Broadwell, or Skylake CPUs~\cite{hazelwood2018applied}.
These servers include large capacity DRAMs and support wide-SIMD instructions that are used for running memory and compute intensive ML inference jobs.

Table \ref{tab:machines} describes the key architecture features of the Intel CPU server systems considered in this paper.
Compared to Skylake, Haswell and Broadwell servers have higher operating frequencies.
For consistency, turbo boost is disabled for all experiments in this paper.
On the other hand, the Skylake architecture has support for  AVX-512 instructions, more parallel cores, and larger L2 caches.
Furthermore, Haswell and Broadwell implement an inclusive L2/L3 cache hierarchy, while Skylake implements a non-inclusive/exclusive cache-hierarchy~\cite{jaleel2015high, jaleel2010achieving}. 
(For the remainder of this paper we will refer to Skylake's L2/L3 cache hierarchy as exclusive).
Sections~\ref{sec:single}~and~\ref{sec:colocation} describe the tradeoff between the system and micro-architecture designs, and their impact on inference latency and throughput in the data center.

\textbf{Synthetic recommendation models:}
To study the performance characteristics of recommendation models, we consider a representative implementation of the three model types RMC1, RMC2 and RMC3 shown in Table \ref{tab:RMParams}.
We analyze inference performance using a benchmark~\cite{DLRM} which accurately represents the execution flow of production-scale models (Section \ref{sec:opensource}). The benchmark is implemented in Caffe2 with Intel MKL as a backend library. All experiments are run with a single Caffe2 worker and Intel MKL thread.

Inputs and models must be processed in parallel to maximize throughput (i.e., number of posts) processed under strict SLA requirements.
This is accomplished by using non-unit batch-sizes and co-locating models on a single system (see Section \ref{sec:single} and Section \ref{sec:colocation}).
All data and model parameters are stored in \texttt{fp32} format.


\section{Understanding Inference \\ Performance of a Single Model} \label{sec:single}
In this section we analyze the performance of a single production-scale recommendation model running on server class Intel CPU systems.

\begin{figure}[t!]
  \centering
  \includegraphics[width=\columnwidth]{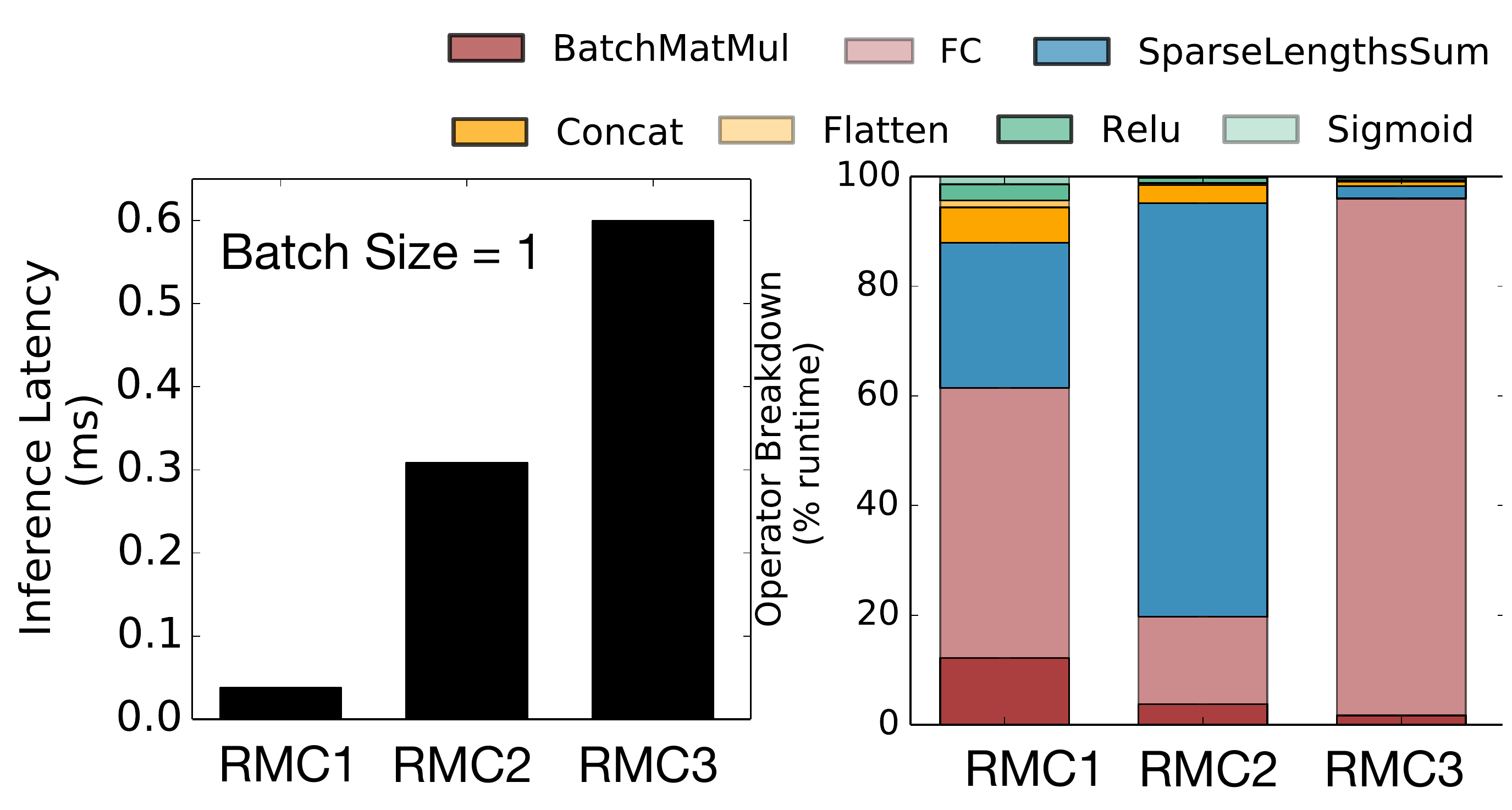}
  \vspace{-1em}
  \caption{(Left) Inference latency of three at-scale recommendation models (RMC1, RMC2, RMC3) on an Intel Broadwell server, unit batch size, varies by an order of magnitude. 
  (Right) Breakdown of time spent, unit batch size, in each operator also varies significantly across the three models.}
  \vspace{-1em}
    \label{fig:single_inference}
\end{figure}

\textbf{Takeaway-message 1:} \textit{Inference latency varies by 15$\times$ across production-scale recommendation models}.

Figure \ref{fig:single_inference} (left) shows the inference latency of the three classes of production-scale models, with unit batch-size, on an Intel Broadwell server.
RMC1 and RMC2 have a latency of 0.04$ms$ and 0.30$ms$, respectively.
This is a consequence of the size of the embedding tables which are an order of magnitude larger in RMC2.
Compared to RMC1 and RMC2, however, RMC3 has a much higher latency of 0.60$ms$.
This is because RMC3 has significantly larger FC layers.
Furthermore, we find significant latency differences between small and large implementations of each type of recommendation model. 
For instance, a large RMC1 has a 2$\times$ longer inference latency as compared to a small RMC1 model, due to more embedding tables and larger FC layers (Table \ref{tab:RMParams}).

\textbf{Takeaway-message 2:}
\textit{While embedding tables set memory requirements, no single operator determines the runtime bottleneck across recommendation models.}

Figure \ref{fig:single_inference} (right) shows the breakdown of execution time for the three classes of production-scale models running on an Intel Broadwell server.
The trends of operator level breakdown across the three recommendation models hold for different Intel server architectures (across Haswell, Broadwell, Skylake).
When running compute intensive recommendation models, such as RMC3, over 96\% of the time is spent in either the BatchMatMul or FC operators.
However, the BatchMatMul and FC operators comprise only 61\% of the run-time for RMC1.
The remainder of the time is consumed by running SparseLengthsSum (20\%), which corresponds to embedding table operations in Caffe2, Concat (6.5\%), and element-wise activation functions.
In contrast, for memory-intensive production-scale recommendation models, like RMC2, SparseLengthsSum consumes 80\% of the execution time of the model.

Thus, software and hardware acceleration of matrix multiplication operations alone (e.g., BatchMatMul and FC) will provide limited benefits on end-to-end performance across all three recommendation models.
Solutions for optimizing the performance of recommendation models  must consider efficient execution of non-compute intensive operations such as embedding table lookups.

\textbf{Takeaway-message 3:} \textit{Running production-scale recommendation models on Intel Broadwell optimizes single model inference latency.}

\begin{figure}[t!]
  \centering
  \includegraphics[width=\columnwidth]{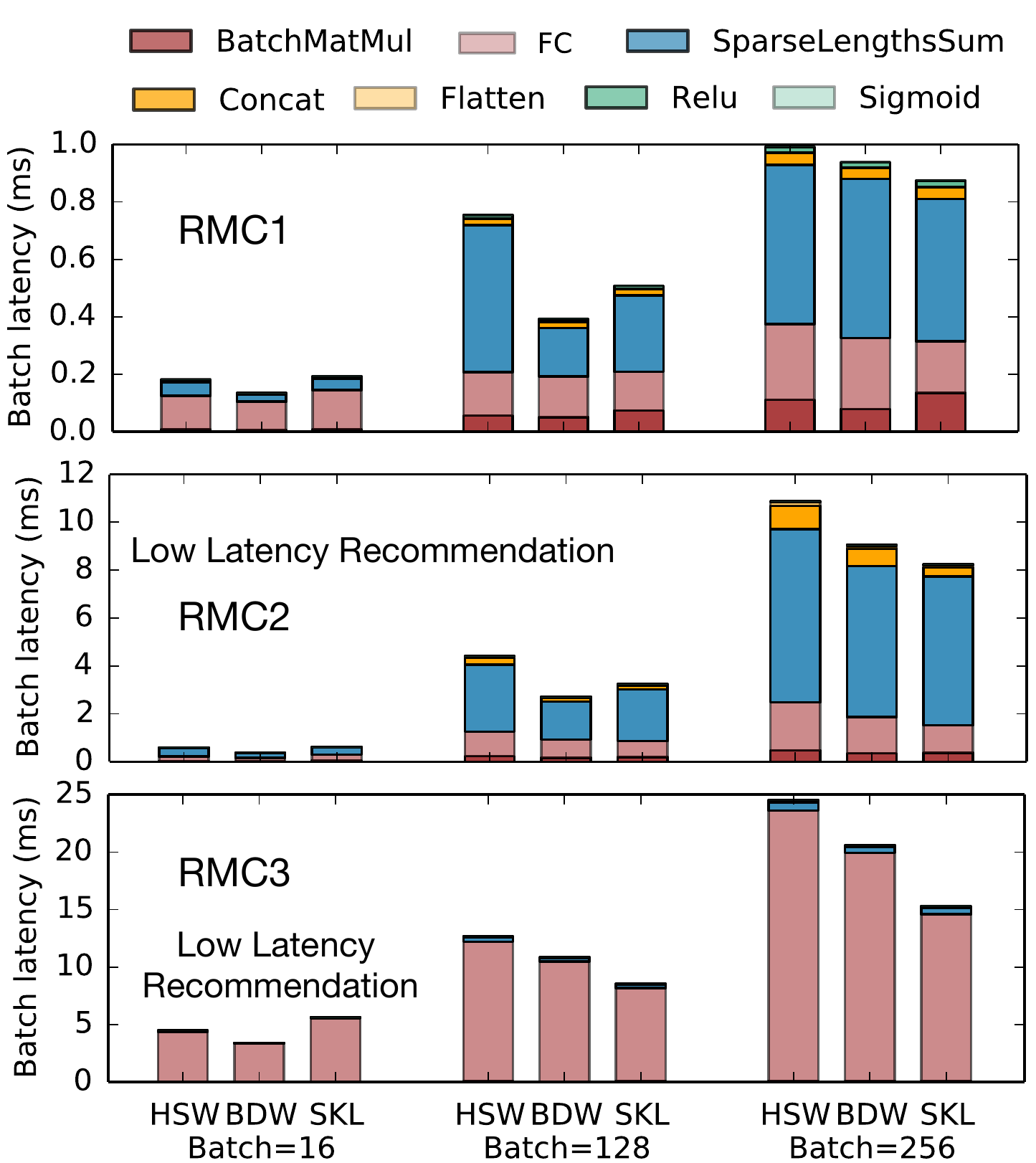}
  \caption{Inference latency of RMC1 (Top), RMC2
(Center), and RMC3 (Bottom) with batch sizes of 16, 128,
and 256. 
While Broadwell is optimal at low batch-sizes, Skylake has higher performance with larger batch-sizes. 
This is a result of Skylake's wider-SIMD (AVX-512) support. 
The horizontal line threshold indicates SLA requirements in low-latency recommendation systems (e.g., search~\cite{tpu, wideanddeep}
).
  }
  \vspace{-1.25em}
  \label{fig:batch_inference}

\end{figure}

Figure \ref{fig:batch_inference} compares the inference latency of running the recommendation models on Intel Haswell, Broadwell, and Skylake servers.
We vary the input batch-size from 16, 128, to 256 for all three recommendation models RMC1(top), RMC2(center), and RMC3(bottom).
For a small batch size of 16, inference latency is optimized when the recommendation models are run on the Broadwell architecture.
For instance, compared to the Haswell and Skylake architectures, Broadwell sees 1.4$\times$ and 1.5$\times$ performance improvement for RMC1, 1.3$\times$ and 1.4$\times$ performance improvement for RMC2, and 1.32$\times$ and 1.65$\times$ performance improvement on RMC3.

At low batch sizes, Broadwell outperforms Skylake due a higher clock frequency.
As shown in Table~\ref{tab:machines}, Broadwell has a 20\% higher clock frequency compared to Skylake.
While Skylake has wider-SIMD support with AVX-512 instructions, recommendation models with smaller batch sizes (e.g., less than 16) are memory bound and do not efficiently exploit the wider-SIMD instruction.
For instance, we can measure the SIMD throughput by measuring the number of fp\_arith\_inst\_retired (512b\_packed\_single) instructions using the Linux \textit{perf} utility.
The SIMD throughput with a batch-size of 4 and 16 are 2.9$\times$ (74\% of theoretical) and 14.5$\times$ (91\% of theoretical) higher, respectively, as compared that with unit batch-size.
As a result, for small batch-sizes Broadwell outperforms Skylake, due to its higher clock frequency and the inefficient use of AVX-512 in Skylake.

Broadwell machines outperform Haswell machines due to a higher DRAM frequency.
Haswell's longer execution time is caused by its slower execution of the SparseLengthsSum operator.
Recall that the SparseLengthSum operator is memory intensive.
For instance, the LLC miss rate of the SparseLengthsSum operator itself is between 1-10 MPKI (see Figure~\ref{fig:embtable}) which corresponds to a DRAM bandwidth utilization of $\sim$ 1GB/s.
As a result, the performance difference between Broadwell and Haswell for the SparseLengthsSum operator comes from differences in DRAM frequency/throughput.
Haswell includes a slower DRAM (DDR3 at 1600MHz) as compared to Broadwell (DDR4 at 2400MHz).

\textbf{Takeaway-message 4:} \textit{While the Skylake has wider-SIMD support, which should provide performance benefits on batched and compute-intensive inference, its throughput is sub-optimal due to irregular memory access patterns from embedding table lookups.}

Recall that in production data centers, recommendation queries for many users and posts must be ranked simultaneously.
One solution to improving overall system throughput is batching.
As shown in Figure \ref{fig:batch_inference}, Skylake exhibits lower run-time with higher batch-sizes.
As a result, for use cases with strict latency constraints (i.e., around 10$ms$ for search~\cite{tpu, wideanddeep}), Skylake can process recommendation with higher batch-sizes.

This is consequence of the Skylake architecture's ability to accelerate FC layers using wider-SIMD support with AVX-512 instructions. 
However, exploiting the benefits of AVX-512 requires much higher batch-sizes, at least 128, for memory intensive production-scale recommendation models, such as RMC1 and RMC2. 
For compute-intensive models, like RMC3, Skylake outperforms both Haswell and Broadwell starting at a batch-size of 64.
These benefits are sub-optimal given Skylake (AVX-512) has a 2$\times$ and 4$\times$ wider SIMD width compared to Broadwell (AVX-2) and Haswell (AVX-2), respectively.
For instance, Skylake runs the memory-intensive RMC1 model 1.3$\times$ faster than Broadwell.
This is due to the irregular memory access patterns from the embedding table lookups.
In fact, the SparseLengthsSum operator becomes the run-time bottleneck in RMC1 with sufficiently high batch-sizes.

\textbf{Takeaway-message 5:} 
\textit{Designers must consider a unique set of performance and resource requirements when accelerating DNN-based recommendation models.} 

First, solutions must balance low-latency, for use cases with stricter SLA (e.g., search \cite{wideanddeep, tpu}), and high-throughout for web-scale services.
Thus, even for inference, hardware solutions \textit{must} consider batching.
This can affect whether performance bottlenecks come from the memory-intensive embedding-table lookups or compute intensive FC layers.
Next, optimizing end-to-end model performance of recommendation workloads requires full-stack optimization given the diverse memory capacity, compute intensity, and memory access pattern characteristics seen in representative implements (e.g., RMC1, RMC2, RMC3).
For instance, a combination of aggressive compression and novel memory technologies~\cite{eisenman2018bandana} are needed to reduce the memory capacity requirements.
%
Existing solutions of standalone FC accelerators~\cite{chen2014dadiannao,tpu, eie, eyeriss, minerva,cambricon-x} will provide limited performance, area, and energy benefits to recommendation models.
Finally, accelerator architectures must balance flexibility with efficient resource utilization, in terms of memory capacity, bandwidth, and FLOPs, to support the diverse set data center use cases.

\begin{figure}[t!]
  \centering
    \includegraphics[width=\columnwidth]{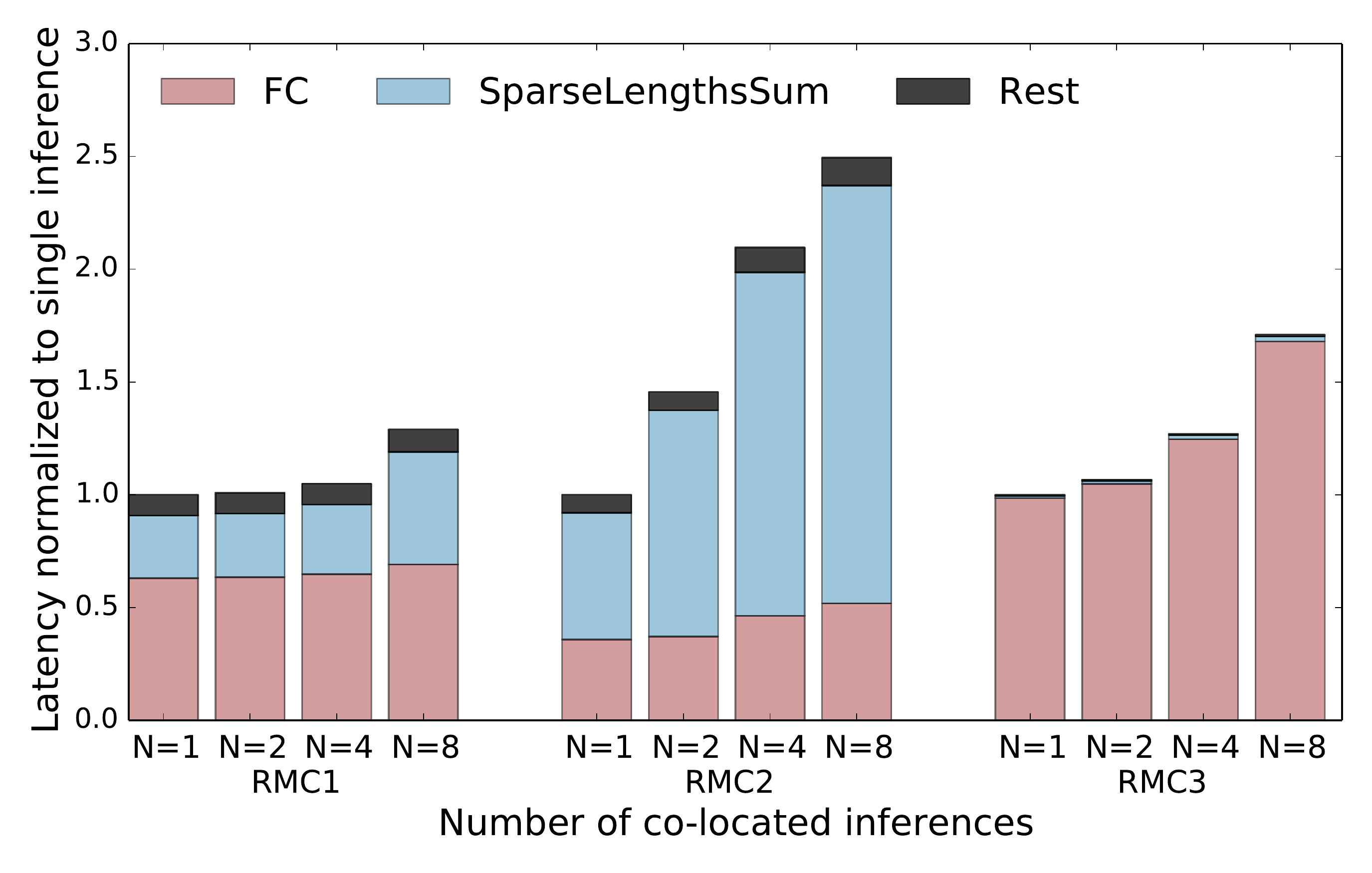}
    \vspace{-1em}
  \caption{Impact of co-locating production-scale recommendation models on Broadwell.
  Increasing the number of co-located models degrades per-model latency.
  RMC2 latency is the most affected by co-location as in FC and SparseLengthsSum run-time degrade by 1.6$\times$ and 3$\times$. }
  \vspace{-1.0em}
  \label{fig:colocatedperf}
\end{figure}

\section{Understanding Effects of \\ Co-locating Models} \label{sec:colocation}
In addition to batching multiple items into a single inference, multiple RM inferences are simultaneously run on the same server in order to service billions of requests world-wide. 
This translates to higher resource utilization.
Co-locating multiple production-scale recommendation models on a single machine can however significantly degrade inference serving latency, trading off single model latency with server  throughput. 

We analyze the impact of co-location on per-model latency as well as overall throughput due to co-location. 
We find that the effects of co-location on latency and throughput depend on not only the type of production-scale recommendation model but also the underlying server architecture. 
For instance, processor architectures with inclusive L2/L3 cache hierarchies (i.e., Haswell, Broadwell) are particularly susceptible to performance degradation and increased performance variability, compared to processors with exclusive L2/L3 cache hierarchies (i.e., Skylake).
This exposes opportunities for request scheduling optimization in the data center ~\cite{mars2013whare,beckmann2013jigsaw}.


\textbf{Takeaway-message 6} 
\textit{Per-model latency degrades due to co-locating many production-scale recommendation models on a single machine. In particular, RMC2's latency degrades more than RMC1 and RMC3 due to a higher degree of irregular memory accesses.}

Figure \ref{fig:colocatedperf} shows the model latency degradation as we co-locate multiple instances of the RMC1, RMC2, and RMC3 on a single machine with a batch-size of 32.
To highlight the relative degradation, latency is normalized to that of running a single instance (N=1) of each recommendation model. 
Compared to RMC1 and RMC3, we find that RMC2 suffers higher latency degradation.
For instance, co-locating 8 production-scale models, degrades latency by 1.3, 2.6, 1.6$\times$ for RMC1, RMC2, and RMC3 respectively.
At the data center scale, this introduces opportunities for optimizing the number of co-located models per machine in order to balance inference latency with overall throughput --- number of items ranked under a strict latency constraint given by the SLA requirements.

Figure \ref{fig:colocatedperf} also shows that latency degradation from co-location is caused by lower FC and SparseLengthsSum performance.
As seen in RMC1 and RMC2, the fraction time spent running SparseLengthsSum increases with higher degrees of co-location.
RMC3 remains dominated by FC layers.
For instance, for RMC2, co-location increases time spent on FC and SparseLengthsSum increases by 1.6$\times$ and 3$\times$, respectively. 
While the time spent on remaining operators, accumulated as "Rest", also increases by a factor of 1.6$\times$, the impact on the overall run-time is marginal. 
Similarly, for RMC1 the fraction of time spent running SparseLengthsSum increases from 15\% to 35\% when running 1 job to 8 jobs.

The greater impact of co-location on SparseLengthsSum is due to the higher degree of irregular memory accesses which, compared to FC, exhibits less cache reuse.
\Rebuttal{For instance, by increasing the number of RMC2 co-located models from 1 to 8, the per. model LLC-MPKI miss rate increases from 0.06 to 0.8.}
Thus, while co-location improves overall throughput of high-end server architecture, it can impact performance bottlenecks when running production-scale recommendation model leading to lower resource utilization.

\textbf{Takeaway-message 7} 
\textit{Processor architectures with inclusive L2/L3 cache hierarchies (i.e., Haswell, Broadwell) are more susceptible to per-model latency degradation as compared to ones with exclusive cache hierarchies (i.e., Skylake) due to a high degree of irregular memory accesses in production recommendation models.}

\begin{figure}[t!]
  \centering
  \includegraphics[width=\columnwidth]{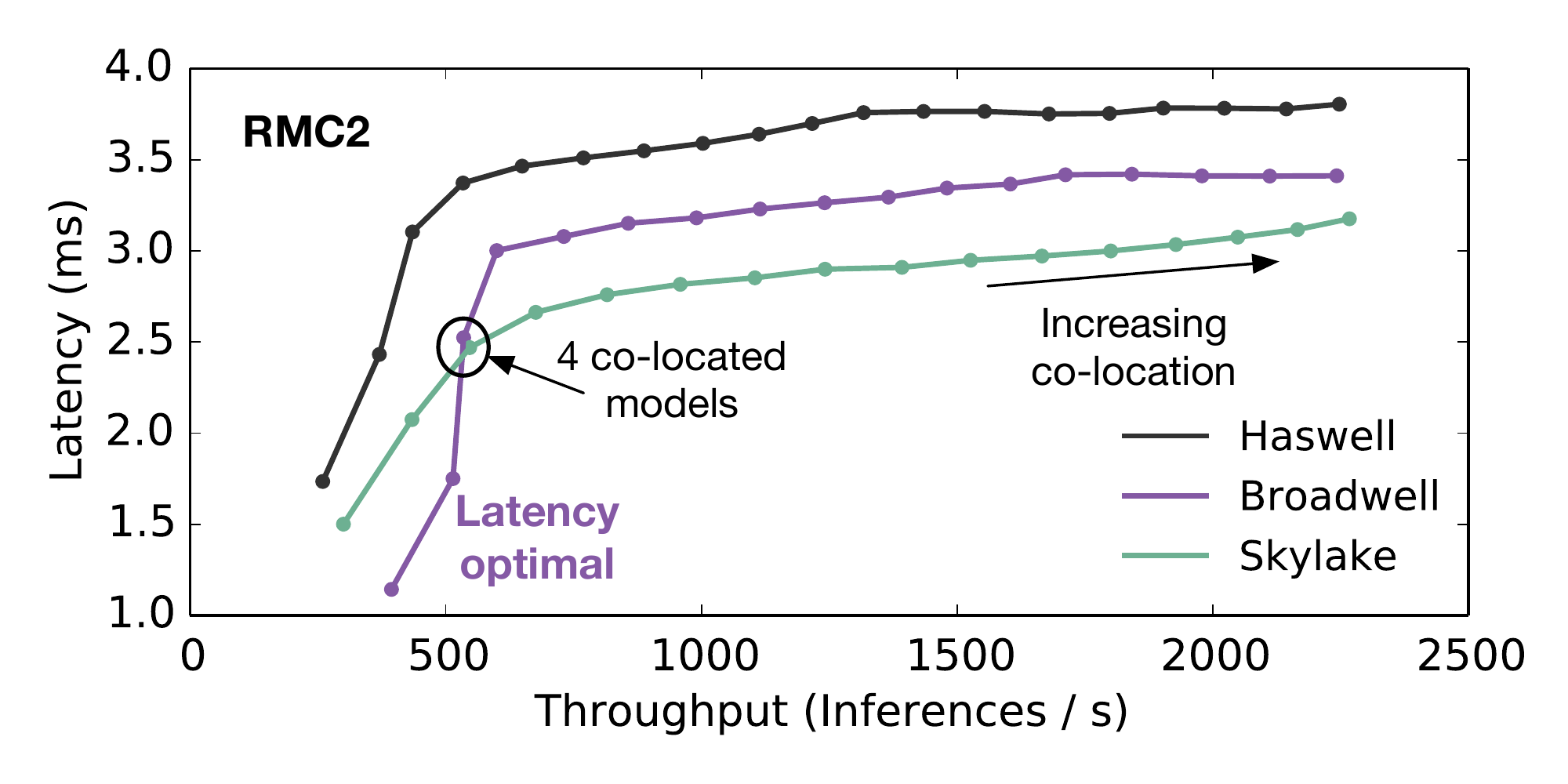}
  \vspace{-1em}
  \caption{Latency/throughput tradeoff with varying number of co-located RMC2 models. 
  Starting from no co-location, latency quickly degrades before plateauing.
  Broadwell performs best under low co-location (latency).
  Skylake is optimal under high co-location (throughput). 
  Skylake's degradation around 18 co-located jobs is due to a sudden increase in LLC miss rate.
  }
  \label{fig:colocatedsys}
  \vspace{-1.0em}
\end{figure}

Figure \ref{fig:colocatedsys} shows the impact of co-locating a production-scale recommendation model on both latency and throughput across the Intel Haswell, Broadwell, and Skylake architectures.
While the results shown are for RMC2, the takeaways hold for RMC1 and RMC3 as well. 
Throughput is measured by the number of inferences per second and bounded by a strict latency constraint, set by the SLA requirement, of 450$ms$.

\textbf{No co-location:}
Recall that in the case of running a single inference per machine, differences in model latency across servers is determined by operating frequency, support for wide-SIMD instructions, and DRAM frequency (see Section~\ref{sec:single} for details).
Similarly, with few co-located inference (i.e., $N=2$), Broadwell has a 10\% higher throughput and lower latency  compared to Skylake.

\begin{figure*}
\centering
        \begin{subfigure}[b]{0.33\textwidth}
                \includegraphics[width=\linewidth]{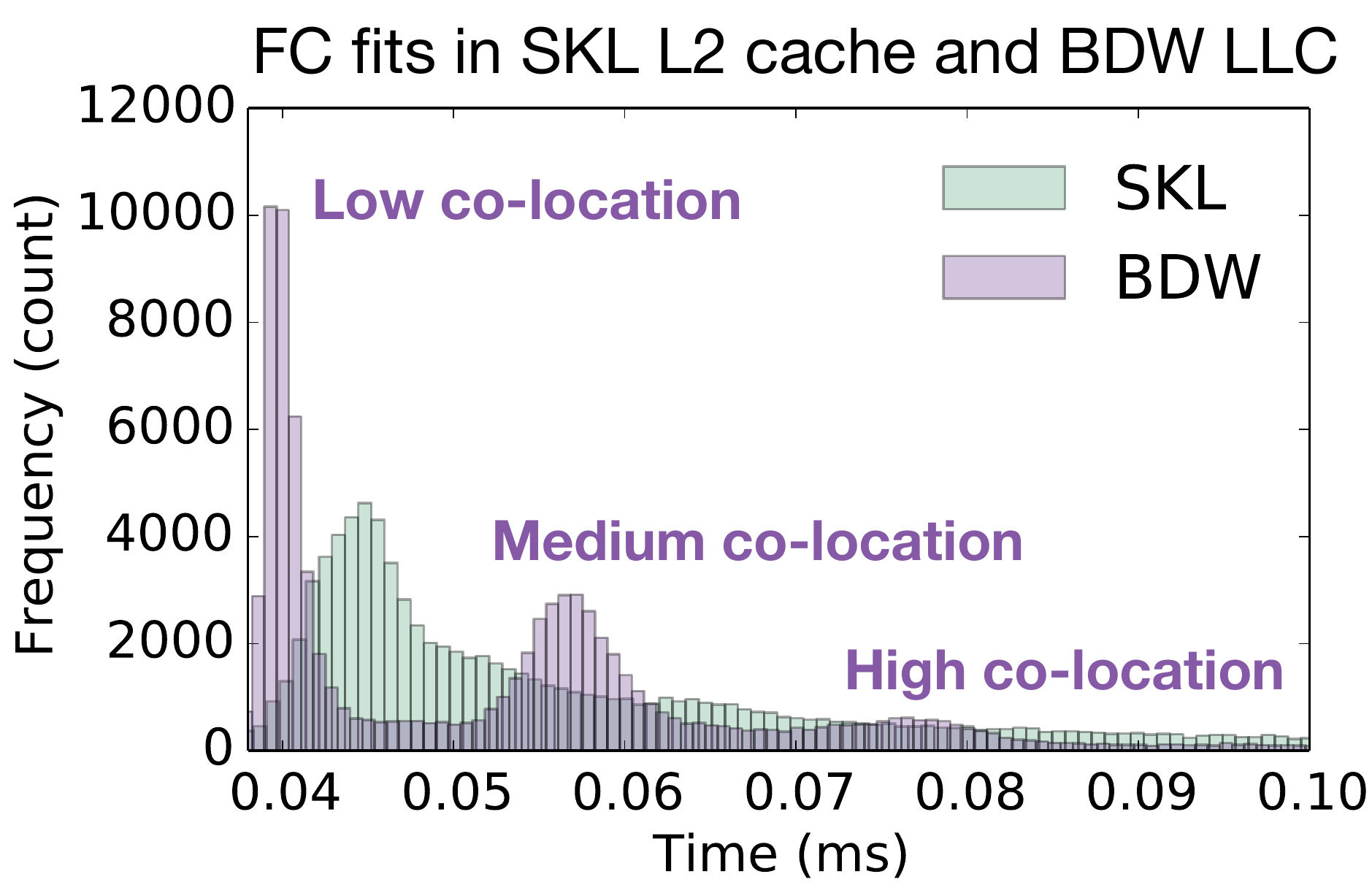}
                \caption{FC in \textit{production-environment}. }
                \label{fig:prod_a}
        \end{subfigure}%
        \begin{subfigure}[b]{0.33\textwidth}
                \includegraphics[width=\linewidth]{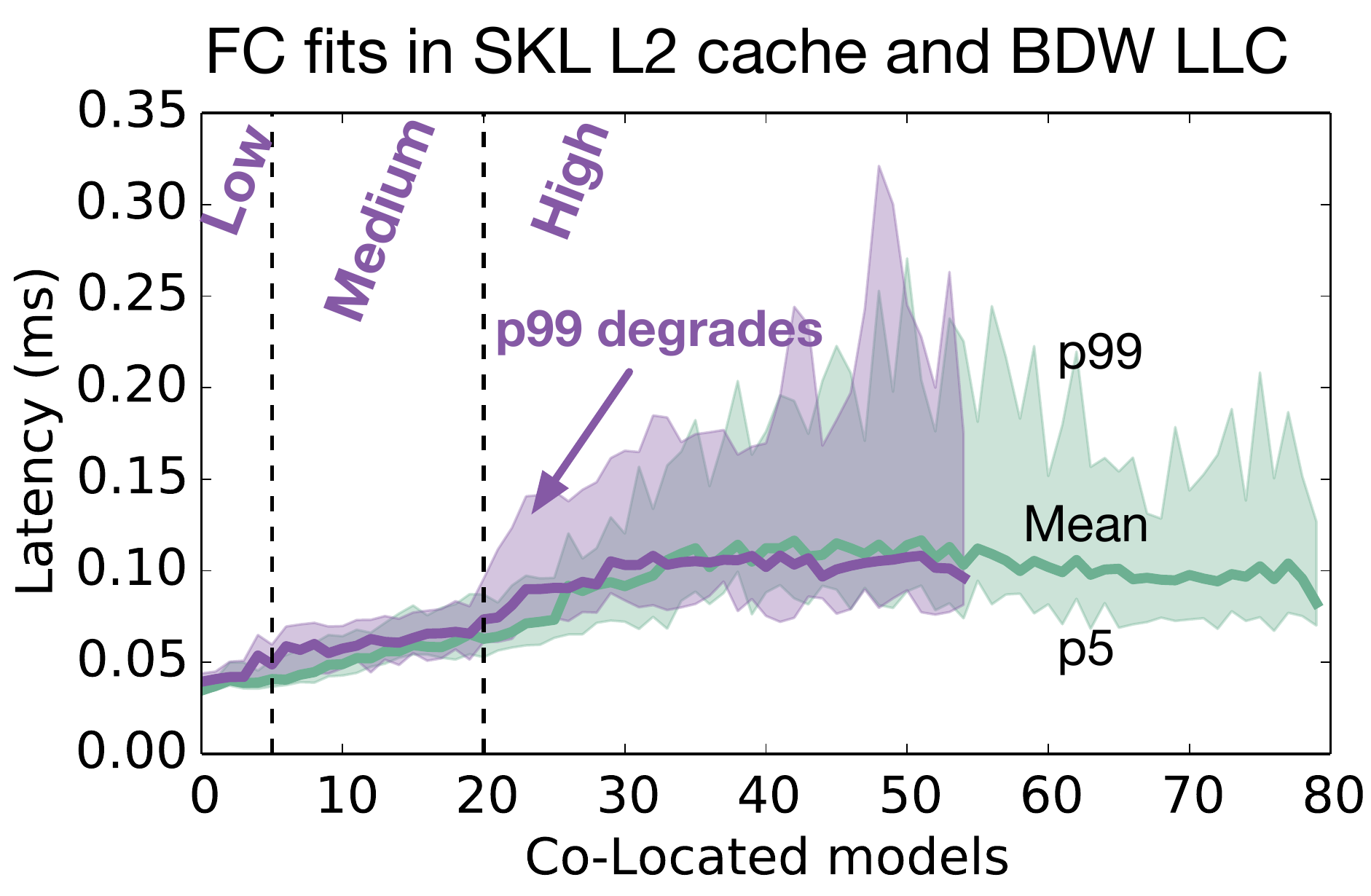}
                \caption{Same FC under co-location}
                \label{fig:prod_b}
        \end{subfigure}%
        \begin{subfigure}[b]{0.33\textwidth}
                \includegraphics[width=\linewidth]{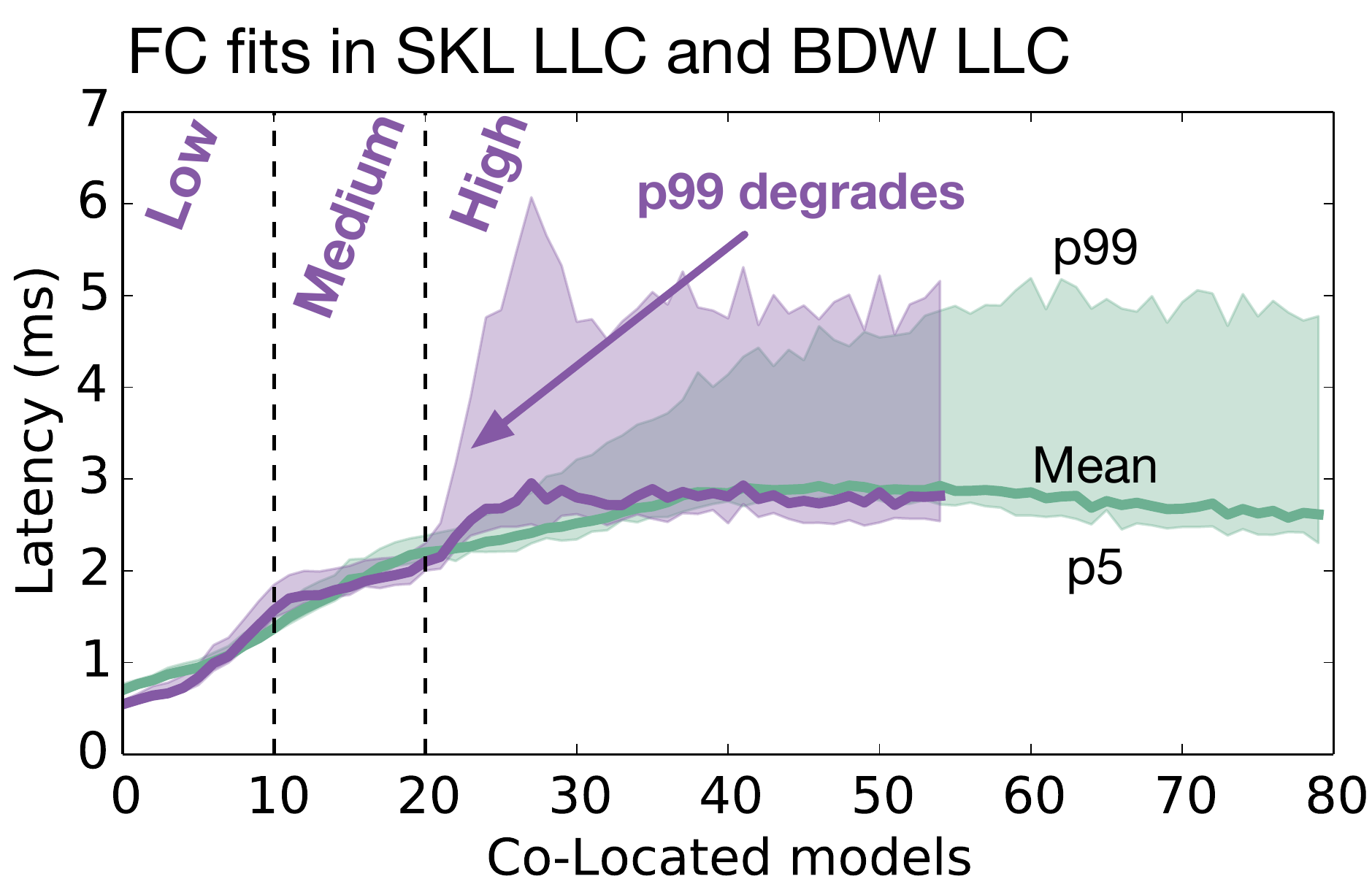}
                \caption{Larger FC under co-location}
                \label{fig:prod_c}
        \end{subfigure}
        \caption{(a) Performance distribution of FC operator that fits in Skylake L2 cache and Broadwell LLC. The three highlighted modes correspond to Broadwell with low, medium, and high co-location. 
        (b) Mean latency of the same FC operator (solid line) increases with more co-location. Gap between p5 and p99 latency (shaded region) increases drastically on Broadwell with high co-location and more gradually on Skylake. 
        (c) Larger FC operator highlights the difference in Broadwell's drastic p99 latency degradation compared to Skylake's gradual degradation. Differences between Broadwell and Skylake under high co-location are due to L2/L3 cache sizes and inclusive/exclusive hierarchies.
        }
        \vspace{-0.5em}
        \label{fig:prod_data}
\end{figure*}

\textbf{Co-locating models:}
Increasing the co-location degree, Skylake outperforms both Haswell and Broadwell in terms of latency and throughput.
Co-locating inferences on a single machine stresses the shared memory system causing latency degradation.
This is particularly true for co-locating production-scale recommendation models that exhibit a high degree of irregular memory accesses.
In contrast, traditional DNNs exhibit higher L1/L2 cache reuse.
Under strict latency bounds (e.g., 3$ms$), Skylake provides the highest throughput by accommodating multiple, co-located recommendation models on a single machine. 

Skylake's higher performance with high co-location is a result of implementing an exclusive L2/L3 cache hierarchy as opposed to an inclusive one.
Inclusive caches suffer from a higher L2 cache miss-rate, due to the irregular memory access patterns in recommendation models.
For instance, Broadwell's L2 miss rate increases by 29\% when running 16 co-located inferences (22 MPKI) compared to a single inference.
Skylake has not only a lower L2 miss rate (13 MPKI for single inference), but also a smaller L2 miss rate increase (10\%).
This is not only caused by a smaller L2 cache size in Broadwell, but also a higher degree of cache back-invalidation due to an inclusive L2/L3 cache hierarchy.
For instance, Broadwell sees a 21\% increase in L2 read-for-ownership miss rate, compared to only 9\% on Skylake. 
Finally, with a high number of co-located inferences (over 18), Skylake suffers from a sudden latency drop caused by a 6\% increase in LLC miss rate.

\textbf{Simultaneous multithreading/hyperthreading}
Prior work has shown that simultaneous multithreading/hyperthreading in modern processors generally improves system throughput~\cite{tullsen1995simultaneous,tullsen1996exploiting}.
However, multithreading/hyperthreading degrades p99 latency for recommendation models, especially  compute-intensive ones (i.e., RMC3).
Results in Figure \ref{fig:colocatedsys} are without hyperthreading --- one model per physical core.
Enabling hyperthreading causes FC and SparseLengthsSum run-times to degrade by 1.6$\times$ and 1.3$\times$, respectively.
The FC operator suffers more performance degradation as it exploits hardware for wide-SIMD instructions (i.e., AVX-2, AVX-512) that are time-shared across threads on the physical core.
As a result, latency degradation due to hyperthreading is more pronounced in compute-intensive recommendation models (i.e., RMC3).


\subsection{Recommendation Inference in Production} \label{sec:production}

The experiments thus far study average latency and throughput across production-scale recommendation models, system architectures, and run-time configurations (e.g., batch-size, number of co-located models).
However, data center execution must also consider tail performance~\cite{kanev2015profiling, dean2013tail}.
Here, we study the impact of co-location on average and tail latency of individual operators.
Production-scale data shows that Broadwell sees a larger performance degradation due to co-location compared to Skylake.

Furthermore, inference for recommendation models running in the data center suffer from high performance variability.
While we do not see performance variability in stand-alone recommendation models (Section~\ref{sec:single} and Section~\ref{sec:colocation}), we find pronounced performance variability for recommendation-models co-located in the production environment.
In fact, p99 latency degrades faster, as the number of co-located inferences increases, on Broadwell machines (inclusive L2/L3 caches) as compared to Skylake.
This exposes opportunities for optimizing data center level scheduling decisions to trade off latency and throughput, with performance variability.   

\textbf{Takeaway-message 8} 
\textit{While co-locating production scale recommendation models with irregular memory accesses increases the overall throughput, it introduces significant performance variability.}

As an example of performance variability in recommendation systems in production environments, Figure \ref{fig:prod_a} shows the distribution of a single FC (input and output dim of 512) operator found in all three types of recommendation models (i.e., RMC1, RMC2, and RMC3).
The production environment has a number of differences compared to co-locating inferences using the synthetic model implementation, including a job scheduler that implements a thread pool with separate queuing model. 
Despite fixed input and output dimensions, performance varies significantly across Broadwell and Skylake architectures.
In particular, Skylake sees a single mode (45$\mu s$) whereas, Broadwell follows a multi-modal distribution (40$\mu s$, 58$\mu s$, and 75$\mu s$) --- a result of co-locating inferences.

Figure \ref{fig:prod_b} shows the impact on latency for the same FC operator under varying degrees of co-located inferences on Broadwell and Skylake in the production data-center environment.
All inferences co-locate the FC operator with RMC1 inferences.
Inferences are first co-located to separate physical cores (i.e., 24 for Broadwell, 40 for Skylake) and exploit then hyper-threading.
The solid lines illustrate the average operator latency on Broadwell (red) and Skylake (blue), while the shaded regions represent the p5 (bottom) and p99 (top) latencies.

Three key observations are made here.
First, average latency increases with more co-location.
On Broadwell the average latency of the FC operator can be categorized into three regions: 40$\mu s$ (no co-location), 60$\mu s$ (5-15 co-located jobs), and 100$\mu s$ (over 20 co-located jobs).
This roughly corresponds to the modes seen in Figure \ref{fig:prod_a}.
Second, the p99 latency increases significantly with high co-location (over 20 jobs) on Broadwell. 
Thus, increasing average throughput with co-location sacrifices predictably meeting SLA requirements.
Third, the average and p99 latency increases more gradually on Skylake.
This is a result of an exclusive L2/L3 cache-hierarchy in Skylake --- the impact of co-locating recommendation models with irregular memory accesses is less on the shared memory system.

Figure \ref{fig:prod_c} runs the same experiments for a much larger FC operator to highlight the key observations: (1) three regions (no-location, 10-15 co-located jobs, more than 20 co-located jobs) of operator latency on Broadwell, (2) large increase in p99 latency under high co-location, and (3) gradual latency degradation on Skylake.
Broadwell suffers from higher performance variability as compared to Skylake.
This exposes opportunities for scheduling optimizations to balance latency, throughput, and performance variability.

\section{Open-Source Benchmark}\label{sec:opensource}

\begin{figure}[t!]
  \centering
  \includegraphics[width=\columnwidth]{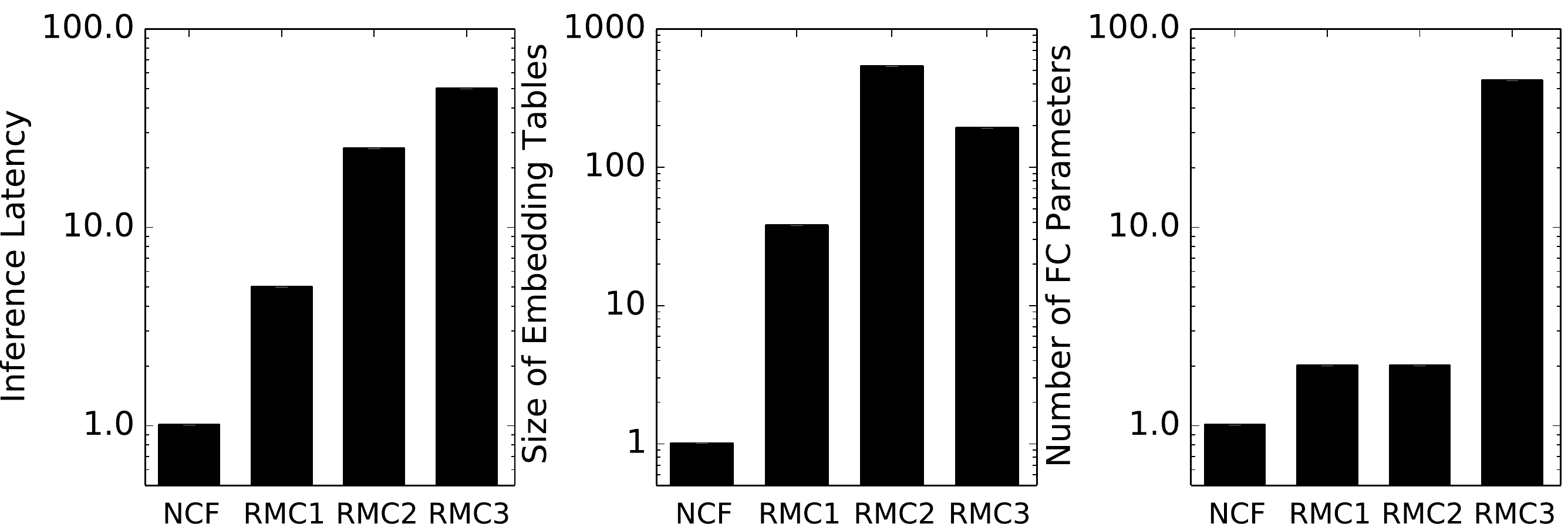}
  \vspace{-1em}
  \caption{ At-scale recommendation models (RMC1, RMC2, RMC3)
have orders of magnitude longer inference latency, larger embedding tables, and FC layers compared to MLPerf-NCF. 
All parameters are normalized to MLPerf-NCF.}
  \label{fig:motiv}
  \vspace{-0.5em}
\end{figure}

Publicly available DNN based recommendation benchmarks, i.e., neural-collaborative filtering (MLPerf-NCF~\cite{mlperf,ncf}), are not representative of the ones used in the data center. 
For instance, compared to production-scale recommendation workloads, the NCF workload from MLPerf~\cite{mlperf} has orders of magnitude smaller embedding tables and fewer FC parameters (Figure~\ref{fig:motiv}). Consequently, FC comprises over 90\% of the execution time in NCF, in contrast SparseLengthSum comprises around 80\% of the cycles in RMC1 (with batching) and RMC2. 
This section describes an open-source benchmark that represents data center scale implementations of Facebook's DNN-based recommendation models~\cite{DLRM}.
The goal is to close the gap between currently available and realistic production-scale benchmarks. 




\subsection{Configuring the open-source benchmark}
\begin{figure}[t!]
  \centering
  \includegraphics[width=\columnwidth]{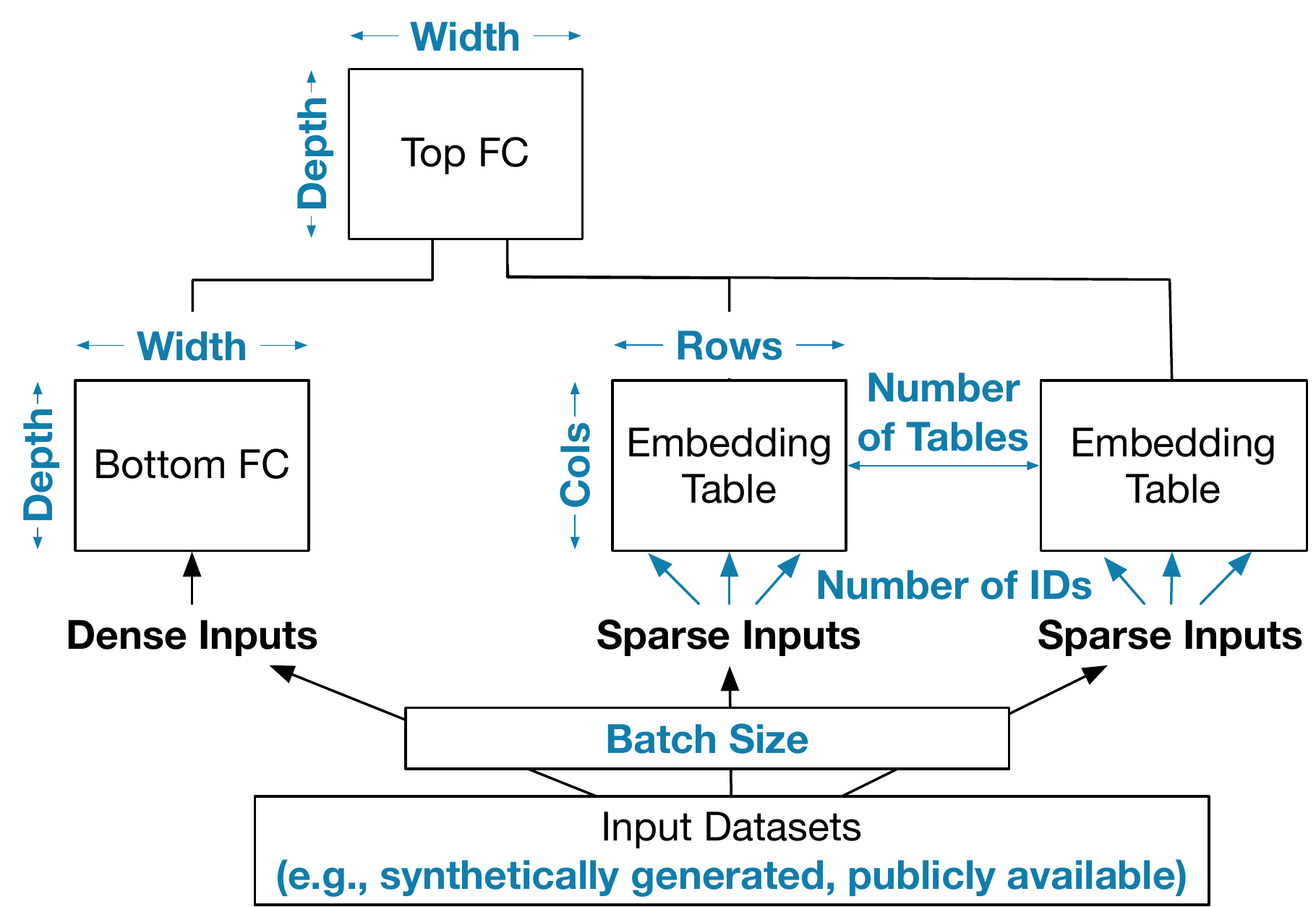}
  \caption{Overall architecture of the open-source recommendation model system. 
  All configurable parameters are outlined in blue.}
  \label{fig:opensource}
   \vspace{-0.5em}
\end{figure}

The open-source benchmark~\cite{DLRM,FBDLRM} was designed with the flexibility to not only study the production scale models seen in this paper (i.e., RMC1, RMC2, RMC3), but also a wider set of realistic recommendation models (e.g., personalized ranking of video content~\cite{youtube}). 
To facilitate ease of use and maximize flexibility, the open-source benchmark provides a suite of tunable parameters to define an end-to-end recommendation system, as shown in Figure \ref{fig:opensource}. 
The set of configurable parameters include: (1) the number of embedding tables, (2) input and output dimensions of embedding tables, (3) number of sparse lookups per embedding table, (4) depth/width of MLP layers for dense features (Bottom-MLP), and (5) depth/width of MLP layers after combining dense and sparse features (Top-MLP). 
These parameters can be configured to implement recommendation models dominated by dense feature processing (i.e., RMC1, RMC3) and sparse feature processing (i.e., RMC1, RMC2). 
Table~\ref{tab:summary} summarizes the key micro-architectural performance bottlenecks for the different classes of recommendation models studied in this paper.

\textbf{Example configurations:}
As an example, let's consider an RMC1 model (Table \ref{tab:RMParams}). In this model the number of embedding tables can be set to 5, with input and output dimensions of $10^5$ and $32$, the number of sparse lookups to $80$, depth and width of BottomFC layers to $3$ and $128$-$64$-$32$, and the depth and width of TopFC layers to $3$ and $128$-$32$-$1$. 

\textbf{Using the open-source benchmark}
The open-source DLRM benchmark is used to study recommendation models in this paper.
By varying the batch, FC, and embedding table configurations,
it can also be used to study other recommendation models.
More generally, it can been used to analyze scheduling decisions, such as running recommendation models across many nodes (distributed inference) or threads.

\begin{figure}[t!]
  \centering
  \includegraphics[width=\columnwidth]{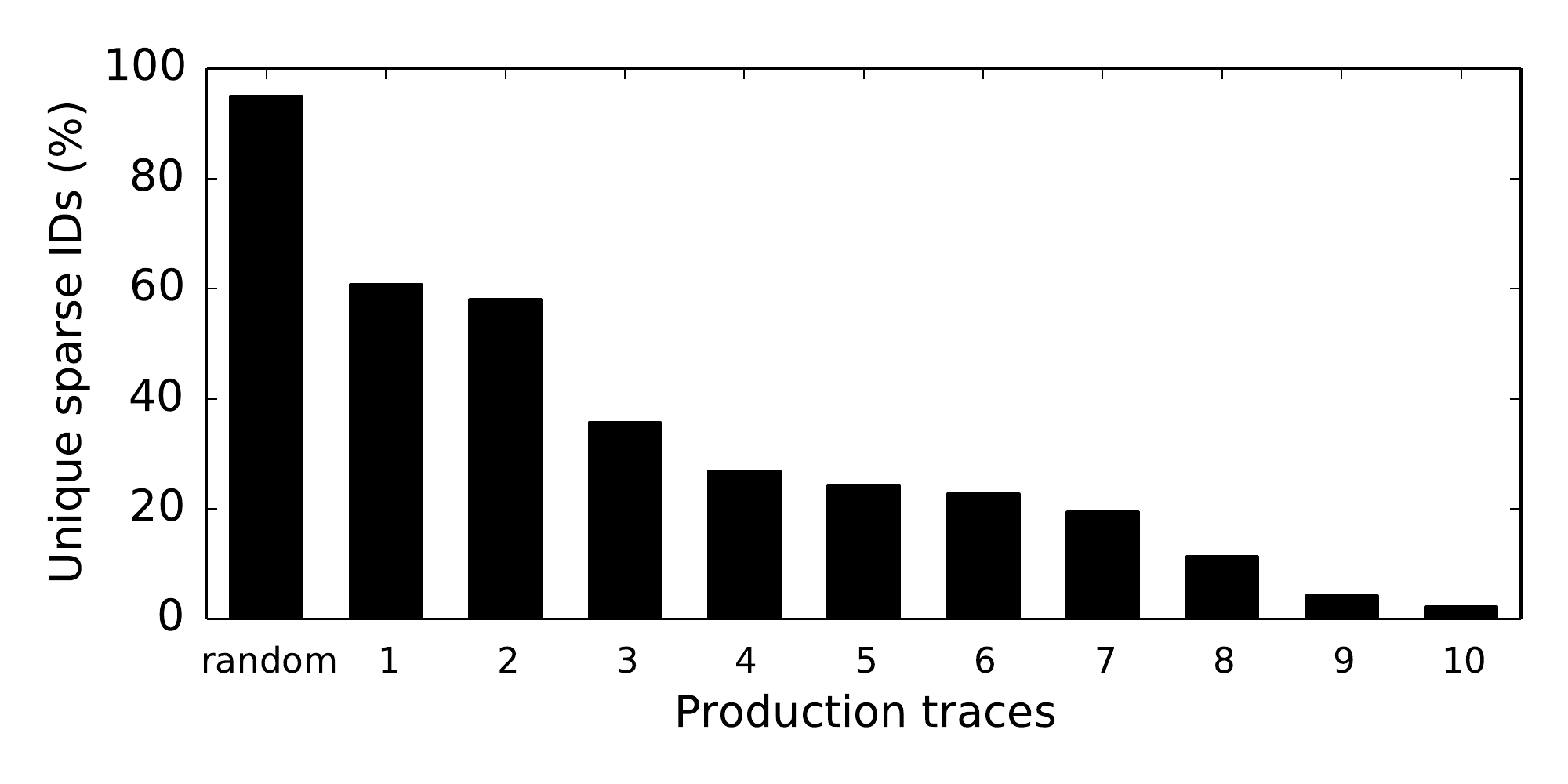}
  \vspace{-1.5em}
  \caption{
  Percent of unique sparse IDs (e.g., embedding table lookups) varies across recommendation use cases and production traces.  
  This enables opportunities for intelligent cache and prefetching optimizations. 
  The open-source implementation provides embedding trace generators in order to instrument recommendation models to study memory system optimizations.
  }
   \label{fig:bandana}
  \vspace{-0.75em}
\end{figure}

Finally, the open-source benchmark can be used to design memory systems, intelligent pre-fetching/caching techniques, and emerging memory technologies.
For instance, while the memory access patterns of recommendation models are irregular compared to well-studied CNNs, the memory accesses are highly dependent on inputs and not completely random. Figure~\ref{fig:bandana} illustrates the fraction of unique sparse IDs used to index embedding tables over a variety of production recommendation use cases.
Intuitively, the degree of unique IDs varies based on user behavior (inputs).
Use cases with fewer unique IDs enable opportunities for embedding vector re-use and intelligent caching.
To study the implications of this locality on memory systems, the recommendation model implementation can be instrumented with open-source data sets~\cite{criteo,movielens} as well as a provided load generator~\cite{DLRM}.

\section{Related Work}
While the systems and computer architecture community has devoted significant efforts to performance analysis and optimization for DNNs, relatively little focus has been devoted to personalized recommendation systems.
This section first reviews DNN-based solutions for personalized recommendation.
This is followed by a discussion on state-of-the-art performance analysis and optimizations for DNNs with context on how the proposed techniques relate to recommendation systems. 

\textbf{DNN-based personalized recommendation}
Compared to image-classification~\cite{resnet}, object detection~\cite{yolo}, and speech recognition~\cite{ds1,ds2,ds3} which process dense features, inputs to personalized recommendation are a mix of both dense and sparse features. 
NCF \cite{ncf} uses a combination of embedding table, FC layers, and ReLU non-linearities using the open-source MovieLens-20m dataset \cite{movielens}.
Dense and sparse features are combined using a series of matrix-factorization and FC layers.
In~\cite{youtube}, the authors discuss applying this model architecture to Youtube video recommendation.
A similar model architecture is applied to predict click-through-rates~\cite{wang2017deep}.
More generally, \cite{wideanddeep} explores the the accuracy tradeoff of wide (few FC layers and embedding tables) and deep (many FC layers and embedding tables) for serving recommendation in the Google Play Store mobile application.
The authors find that accuracy is optimized using a combination of wide and deep neural networks, similar to the production-scale recommendation models considered in this paper.
While on-going research explores using CNNs and RNNs in recommendation systems \cite{open-rec}, for the purposes of this paper we focus on production-scale recommendation models (Figure ~\ref{fig:sparsenn}). 


\textbf{DNN performance analysis and optimization}
Current publicly available benchmarks \cite{coleman2017dawnbench,zhu2018benchmarking,fathom,deepbench} for DNNs focus on models with FC, CNN, and RNN layers only. 
In combination with open-source implementations of state-of-the-art networks in high-level deep learning frameworks~\cite{tensorflow,caffe2,pytorch}, the benchmarks have enabled thorough performance analysis and optimization.
However, the resulting software and hardware solutions~\cite{tpu, minerva, eyeriss, eie, cambricon-x,chen2014dadiannao} do not apply to production-scale recommendation workloads.

\begin{table}[t!]
\begin{center}
\small
\begin{tabular}{|c||c|c|}
\hline
\textbf{} & \textbf{Dense features} & \textbf{Sparse features} \\ \hline
Model(s) & RMC1 \& RMC3 & RMC1 \& RMC2 \\ \hline
Operators & MLP dominated & Embedding dominated \\ \hline
\multirow{4}{*}{$\mu$arch bottleneck} & Core frequency &  Core frequency \\ 
& Core count&  Core count\\ 
& DRAM capacity&  DRAM capacity\\
& SIMD performance&  DRAM freq. \& BW \\
& Cache size&  Cache contention \\ \hline

\end{tabular}
\end{center}
  \caption{ 
  \Camera{Summary of recommendation models and key micro-architectural features that impact at-scale performance.}
  }
  \label{tab:summary}
\end{table}

Recommendation workloads pose unique challenges in terms of memory capacity, irregular memory accesses, diversity in compute intensive and memory intensive models, and high-throughput and low-latency optimization targets.
Furthermore, available implementations of DNN-based recommendation systems (i.e, MLPerf NCF\cite{mlperf}) are not representative of production-scale ones.
To alleviate memory capacity and bandwidth constraints, Eisenman et al. propose storing recommendation-models in  non-volatile-memories with DRAM to cache embedding-table queries~\cite{eisenman2018bandana}.
Recent work has also proposed solutions based on near-memory processing to accelerate embedding table operations~\cite{kwon2019tensordimm,ke2019recnmp}.
The detailed performance analysis in this paper will enable future work to consider a broader set of solutions to optimize end-to-end personalized recommendation systems currently running in data centers and motivate additional optimization techniques that address challenges specifically for mobile~\cite{wu2019fb_inference}.



\section{Conclusion}

This paper provides a detailed performance analysis of recommendation models on server-scale systems present in the data center.
The analysis demonstrates that DNNs for recommendation pose unique challenges to efficient execution as compared to traditional CNNs and RNNs.
In particular, recommendation systems require much larger storage capacity, produce irregular memory accesses, and consist of a diverse set of operator-level performance bottlenecks.
The analysis also shows that based on the performance target (i.e., latency versus throughput) and the recommendation model being run, the optimal platform and run-time configuration varies. 
Furthermore, micro-architectural platform features, such as processor frequency and core count, SIMD width and utilization, cache capacity, inclusive versus exclusive cache hierarchies, and DRAM configurations, expose scheduling optimization opportunities for running recommendation model inference in the data center.

For the servers considered in this paper, Broadwell achieves up to 40\% lower latency while Skylake achieves 30\% higher throughput.
This paper also studies the effect of co-locating inference jobs, as mechanisms to improve resource utilization, on performance variability. 
The detailed performance analysis of production-scale recommendation models lay the foundation for future full-stack hardware solutions targeting personalized recommendation.

\newpage


\bibliographystyle{IEEEtranS}
\bibliography{main}

\begin{thebibliography}{10}
\providecommand{\url}[1]{#1}
\csname url@samestyle\endcsname
\providecommand{\newblock}{\relax}
\providecommand{\bibinfo}[2]{#2}
\providecommand{\BIBentrySTDinterwordspacing}{\spaceskip=0pt\relax}
\providecommand{\BIBentryALTinterwordstretchfactor}{4}
\providecommand{\BIBentryALTinterwordspacing}{\spaceskip=\fontdimen2\font plus
\BIBentryALTinterwordstretchfactor\fontdimen3\font minus
  \fontdimen4\font\relax}
\providecommand{\BIBforeignlanguage}[2]{{%
\expandafter\ifx\csname l@#1\endcsname\relax
\typeout{** WARNING: IEEEtranS.bst: No hyphenation pattern has been}%
\typeout{** loaded for the language `#1'. Using the pattern for}%
\typeout{** the default language instead.}%
\else
\language=\csname l@#1\endcsname
\fi
#2}}
\providecommand{\BIBdecl}{\relax}
\BIBdecl

\bibitem{alibabaRec}
\BIBentryALTinterwordspacing
``Breakthroughs in matching and recommendation algorithms by alibaba.''
  [Online]. Available:
  \url{https://www.alibabacloud.com/blog/breakthroughs-in-matching-and-recommendation-algorithms-by-alibaba_593976}
\BIBentrySTDinterwordspacing

\bibitem{caffe2}
\BIBentryALTinterwordspacing
``Caffe2.'' [Online]. Available: \url{https://caffe2.ai///}
\BIBentrySTDinterwordspacing

\bibitem{criteo}
\BIBentryALTinterwordspacing
``Crieto dataset.'' [Online]. Available:
  \url{https://labs.criteo.com/2014/02/download-dataset/}
\BIBentrySTDinterwordspacing

\bibitem{deepbench}
\BIBentryALTinterwordspacing
``Deep bench.'' [Online]. Available: \url{https://deepbench.io///}
\BIBentrySTDinterwordspacing

\bibitem{mlperf}
\BIBentryALTinterwordspacing
``Mlperf.'' [Online]. Available: \url{https://mlperf.org/}
\BIBentrySTDinterwordspacing

\bibitem{movielens}
\BIBentryALTinterwordspacing
``Movielens 20m dataset.'' [Online]. Available:
  \url{https://grouplens.org/datasets/movielens/20m/}
\BIBentrySTDinterwordspacing

\bibitem{pytorch}
\BIBentryALTinterwordspacing
``Pytorch.'' [Online]. Available: \url{https://pytorch.org//}
\BIBentrySTDinterwordspacing

\bibitem{tensorflow}
M.~Abadi, P.~Barham, J.~Chen, Z.~Chen, A.~Davis, J.~Dean, M.~Devin,
  S.~Ghemawat, G.~Irving, M.~Isard \emph{et~al.}, ``Tensorflow: A system for
  large-scale machine learning,'' in \emph{Proceedings of the 12th USENIX
  Symposium on Operating Systems Design and Implementation ($OSDI$ 16)}, 2016,
  pp. 265--283.

\bibitem{fathom}
R.~Adolf, S.~Rama, B.~Reagen, G.-Y. Wei, and D.~Brooks, ``Fathom: Reference
  workloads for modern deep learning methods,'' in \emph{Proceedings of the
  IEEE International Symposium on Workload Characterization (IISWC)}.\hskip 1em
  plus 0.5em minus 0.4em\relax IEEE, 2016, pp. 1--10.

\bibitem{ds2}
D.~Amodei, S.~Ananthanarayanan, R.~Anubhai, J.~Bai, E.~Battenberg, C.~Case,
  J.~Casper, B.~Catanzaro, Q.~Cheng, G.~Chen \emph{et~al.}, ``Deep speech 2:
  End-to-end speech recognition in english and mandarin,'' in \emph{Proceedings
  of the International conference on machine learning}, 2016, pp. 173--182.

\bibitem{ds3}
E.~Battenberg, J.~Chen, R.~Child, A.~Coates, Y.~G.~Y. Li, H.~Liu, S.~Satheesh,
  A.~Sriram, and Z.~Zhu, ``Exploring neural transducers for end-to-end speech
  recognition,'' in \emph{Proceedings of the IEEE Automatic Speech Recognition
  and Understanding Workshop (ASRU)}.\hskip 1em plus 0.5em minus 0.4em\relax
  IEEE, 2017, pp. 206--213.

\bibitem{beckmann2013jigsaw}
N.~Beckmann and D.~Sanchez, ``Jigsaw: Scalable software-defined caches,'' in
  \emph{Proceedings of the 22nd international conference on Parallel
  architectures and compilation techniques}.\hskip 1em plus 0.5em minus
  0.4em\relax IEEE, 2013, pp. 213--224.

\bibitem{instagramSparseNN}
\BIBentryALTinterwordspacing
T.~Bredillet, ``Lessons learned at instagram stories and feed machine
  learning.'' [Online]. Available:
  \url{https://instagram-engineering.com/lessons-learned-at-instagram-stories-and-feed-machine-learning-54f3aaa09e56}
\BIBentrySTDinterwordspacing

\bibitem{eyeriss}
Y.-H. Chen, T.~Krishna, J.~S. Emer, and V.~Sze, ``{Eyeriss: An}
  energy-efficient reconfigurable accelerator for deep convolutional neural
  networks,'' \emph{IEEE Journal of Solid-State Circuits}, vol.~52, no.~1, pp.
  127--138, 2017.

\bibitem{chen2014dadiannao}
Y.~Chen, T.~Luo, S.~Liu, S.~Zhang, L.~He, J.~Wang, L.~Li, T.~Chen, Z.~Xu,
  N.~Sun \emph{et~al.}, ``Dadiannao: A machine-learning supercomputer,'' in
  \emph{Proceedings of the 47th Annual IEEE/ACM International Symposium on
  Microarchitecture}.\hskip 1em plus 0.5em minus 0.4em\relax IEEE Computer
  Society, 2014, pp. 609--622.

\bibitem{wideanddeep}
H.-T. Cheng, L.~Koc, J.~Harmsen, T.~Shaked, T.~Chandra, H.~Aradhye,
  G.~Anderson, G.~Corrado, W.~Chai, M.~Ispir \emph{et~al.}, ``Wide \& deep
  learning for recommender systems,'' in \emph{Proceedings of the 1st Workshop
  on Deep Learning for Recommender Systems}.\hskip 1em plus 0.5em minus
  0.4em\relax ACM, 2016, pp. 7--10.

\bibitem{chui2018notes}
M.~Chui, J.~Manyika, M.~Miremadi, N.~Henke, R.~Chung, P.~Nel, and S.~Malhotra,
  ``Notes from the {AI} frontier insights from hundreds of use cases,'' 2018.

\bibitem{chung2018serving}
E.~Chung, J.~Fowers, K.~Ovtcharov, M.~Papamichael, A.~Caulfield, T.~Massengill,
  M.~Liu, D.~Lo, S.~Alkalay, M.~Haselman \emph{et~al.}, ``Serving {DNNs} in
  real time at datacenter scale with project brainwave,'' \emph{IEEE Micro},
  vol.~38, no.~2, pp. 8--20, 2018.

\bibitem{chung2014empirical}
J.~Chung, C.~Gulcehre, K.~Cho, and Y.~Bengio, ``Empirical evaluation of gated
  recurrent neural networks on sequence modeling,'' \emph{arXiv preprint
  arXiv:1412.3555}, 2014.

\bibitem{coleman2017dawnbench}
C.~Coleman, D.~Narayanan, D.~Kang, T.~Zhao, J.~Zhang, L.~Nardi, P.~Bailis,
  K.~Olukotun, C.~R{\'e}, and M.~Zaharia, ``Dawnbench: An end-to-end deep
  learning benchmark and competition.''

\bibitem{courbariaux2016binarized}
M.~Courbariaux, I.~Hubara, D.~Soudry, R.~El-Yaniv, and Y.~Bengio, ``Binarized
  neural networks: Training deep neural networks with weights and activations
  constrained to+ 1 or-1,'' \emph{arXiv preprint arXiv:1602.02830}, 2016.

\bibitem{youtube}
P.~Covington, J.~Adams, and E.~Sargin, ``Deep neural networks for youtube
  recommendations,'' in \emph{Proceedings of the 10th ACM conference on
  recommender systems}.\hskip 1em plus 0.5em minus 0.4em\relax ACM, 2016, pp.
  191--198.

\bibitem{dean2013tail}
J.~Dean and L.~A. Barroso, ``The tail at scale,'' \emph{Communications of the
  ACM}, vol.~56, no.~2, pp. 74--80, 2013.

\bibitem{Ebrahimi:MICRO09}
E.~Ebrahimi, O.~Mutlu, C.~J. Lee, and Y.~N. Patt, ``Coordinated control of
  multiple prefetchers in multi-core systems,'' in \emph{Proceedings of the
  42Nd Annual IEEE/ACM International Symposium on Microarchitecture}, ser.
  MICRO 42, 2009, pp. 316--326.

\bibitem{eisenman2018bandana}
A.~Eisenman, M.~Naumov, D.~Gardner, M.~Smelyanskiy, S.~Pupyrev, K.~Hazelwood,
  A.~Cidon, and S.~Katti, ``Bandana: Using non-volatile memory for storing deep
  learning models,'' \emph{arXiv preprint arXiv:1811.05922}, 2018.

\bibitem{netflix}
\BIBentryALTinterwordspacing
C.~A. Gomez-Uribe and N.~Hunt, ``The netflix recommender system: Algorithms,
  business value, and innovation,'' \emph{ACM Trans. Manage. Inf. Syst.},
  vol.~6, no.~4, pp. 13:1--13:19, Dec. 2015. [Online]. Available:
  \url{http://doi.acm.org/10.1145/2843948}
\BIBentrySTDinterwordspacing

\bibitem{goyal2017accurate}
P.~Goyal, P.~Doll{\'a}r, R.~Girshick, P.~Noordhuis, L.~Wesolowski, A.~Kyrola,
  A.~Tulloch, Y.~Jia, and K.~He, ``Accurate, large minibatch sgd: Training
  imagenet in 1 hour,'' \emph{arXiv preprint arXiv:1706.02677}, 2017.

\bibitem{gupta2015deep}
S.~Gupta, A.~Agrawal, K.~Gopalakrishnan, and P.~Narayanan, ``Deep learning with
  limited numerical precision,'' in \emph{Proceedings of the International
  Conference on Machine Learning}, 2015, pp. 1737--1746.

\bibitem{eie}
S.~Han, X.~Liu, H.~Mao, J.~Pu, A.~Pedram, M.~A. Horowitz, and W.~J. Dally,
  ``{EIE: E}fficient inference engine on compressed deep neural network,'' in
  \emph{Proceedings of the ACM/IEEE 43rd Annual International Symposium on
  Computer Architecture (ISCA)}.\hskip 1em plus 0.5em minus 0.4em\relax IEEE,
  2016, pp. 243--254.

\bibitem{han2015deep}
S.~Han, H.~Mao, and W.~J. Dally, ``Deep compression: Compressing deep neural
  networks with pruning, trained quantization and huffman coding,'' \emph{arXiv
  preprint arXiv:1510.00149}, 2015.

\bibitem{ds1}
A.~Hannun, C.~Case, J.~Casper, B.~Catanzaro, G.~Diamos, E.~Elsen, R.~Prenger,
  S.~Satheesh, S.~Sengupta, A.~Coates \emph{et~al.}, ``Deep speech: Scaling up
  end-to-end speech recognition,'' \emph{arXiv preprint arXiv:1412.5567}, 2014.

\bibitem{hazelwood2018applied}
K.~Hazelwood, S.~Bird, D.~Brooks, S.~Chintala, U.~Diril, D.~Dzhulgakov,
  M.~Fawzy, B.~Jia, Y.~Jia, A.~Kalro \emph{et~al.}, ``Applied machine learning
  at {Facebook}: a datacenter infrastructure perspective,'' in
  \emph{Proceedings of the IEEE International Symposium on High Performance
  Computer Architecture (HPCA)}.\hskip 1em plus 0.5em minus 0.4em\relax IEEE,
  2018, pp. 620--629.

\bibitem{resnet}
K.~He, X.~Zhang, S.~Ren, and J.~Sun, ``Deep residual learning for image
  recognition,'' in \emph{Proceedings of the IEEE conference on computer vision
  and pattern recognition}, 2016, pp. 770--778.

\bibitem{ncf}
X.~He, L.~Liao, H.~Zhang, L.~Nie, X.~Hu, and T.-S. Chua, ``Neural collaborative
  filtering,'' in \emph{Proceedings of the 26th International Conference on
  World Wide Web}.\hskip 1em plus 0.5em minus 0.4em\relax International World
  Wide Web Conferences Steering Committee, 2017, pp. 173--182.

\bibitem{howard2017mobilenets}
A.~G. Howard, M.~Zhu, B.~Chen, D.~Kalenichenko, W.~Wang, T.~Weyand,
  M.~Andreetto, and H.~Adam, ``Mobilenets: Efficient convolutional neural
  networks for mobile vision applications,'' \emph{arXiv preprint
  arXiv:1704.04861}, 2017.

\bibitem{jaleel2010achieving}
A.~Jaleel, E.~Borch, M.~Bhandaru, S.~C. Steely~Jr, and J.~Emer, ``Achieving
  non-inclusive cache performance with inclusive caches: Temporal locality
  aware (tla) cache management policies,'' in \emph{Proceedings of the 43rd
  Annual IEEE/ACM International Symposium on Microarchitecture}.\hskip 1em plus
  0.5em minus 0.4em\relax IEEE Computer Society, 2010, pp. 151--162.

\bibitem{jaleel2015high}
A.~Jaleel, J.~Nuzman, A.~Moga, S.~C. Steely, and J.~Emer, ``High performing
  cache hierarchies for server workloads: Relaxing inclusion to capture the
  latency benefits of exclusive caches,'' in \emph{Proceedings of the IEEE 21st
  International Symposium on High Performance Computer Architecture
  (HPCA)}.\hskip 1em plus 0.5em minus 0.4em\relax IEEE, 2015, pp. 343--353.

\bibitem{tpu}
N.~P. Jouppi, C.~Young, N.~Patil, D.~Patterson, G.~Agrawal, R.~Bajwa, S.~Bates,
  S.~Bhatia, N.~Boden, A.~Borchers \emph{et~al.}, ``In-datacenter performance
  analysis of a tensor processing unit,'' in \emph{Proceedings of the ACM/IEEE
  44th Annual International Symposium on Computer Architecture (ISCA)}.\hskip
  1em plus 0.5em minus 0.4em\relax IEEE, 2017, pp. 1--12.

\bibitem{judd2016stripes}
P.~Judd, J.~Albericio, T.~Hetherington, T.~M. Aamodt, and A.~Moshovos,
  ``Stripes: Bit-serial deep neural network computing,'' in \emph{Proceedings
  of the 49th Annual IEEE/ACM International Symposium on Microarchitecture
  (MICRO)}.\hskip 1em plus 0.5em minus 0.4em\relax IEEE, 2016, pp. 1--12.

\bibitem{kanev2015profiling}
S.~Kanev, J.~P. Darago, K.~Hazelwood, P.~Ranganathan, T.~Moseley, G.-Y. Wei,
  and D.~Brooks, ``Profiling a warehouse-scale computer,'' in \emph{Proceedings
  of the ACM SIGARCH Computer Architecture News}, vol.~43, no.~3.\hskip 1em
  plus 0.5em minus 0.4em\relax ACM, 2015, pp. 158--169.

\bibitem{ke2019recnmp}
\BIBentryALTinterwordspacing
L.~Ke, U.~Gupta, C.-J. Wu, B.~Y. Cho, M.~Hempstead, B.~Reagen, X.~Zhang,
  D.~Brooks, V.~Chandra, U.~Diril, A.~Firoozshahian, K.~Hazelwood, B.~Jia,
  H.-H.~S. Lee, M.~Li, B.~Maher, D.~Mudigere, M.~Naumov, M.~Schatz,
  M.~Smelyanskiy, and X.~Wang, ``Recnmp: Accelerating personalized
  recommendation with near-memory processing,'' 2019. [Online]. Available:
  \url{https://arxiv.org/abs/1912.12953}
\BIBentrySTDinterwordspacing

\bibitem{kwon2019tensordimm}
Y.~Kwon, Y.~Lee, and M.~Rhu, ``{TensorDIMM: A} practical near-memory processing
  architecture for embeddings and tensor operations in deep learning,'' in
  \emph{Proceedings of the 52nd Annual IEEE/ACM International Symposium on
  Microarchitecture}.\hskip 1em plus 0.5em minus 0.4em\relax ACM, 2019, pp.
  740--753.

\bibitem{mars2013whare}
J.~Mars and L.~Tang, ``Whare-map: Heterogeneity in homogeneous warehouse-scale
  computers,'' in \emph{Proceedings of the ACM SIGARCH Computer Architecture
  News}, vol.~41, no.~3.\hskip 1em plus 0.5em minus 0.4em\relax ACM, 2013, pp.
  619--630.

\bibitem{mlperf-training}
P.~Mattson, C.~Cheng, C.~Coleman, G.~Diamos, P.~Micikevicius, D.~Patterson,
  H.~Tang, G.-Y. Wei, P.~Bailis, V.~Bittorf, D.~Brooks, D.~Chen, D.~Dutta,
  U.~Gupta, K.~Hazelwood, A.~Hock, X.~Huang, B.~Jia, D.~Kang, D.~Kanter,
  N.~Kumar, J.~Liao, D.~Narayanan, T.~Oguntebi, G.~Pekhimenko, L.~Pentecost,
  V.~J. Reddi, T.~Robie, T.~S. John, C.-J. Wu, L.~Xu, C.~Young, and M.~Zaharia,
  ``Mlperf training benchmark,'' \emph{arXiv preprint arXiv:1910.01500}, 2019.

\bibitem{naumov2019dimensionality}
M.~Naumov, ``On the dimensionality of embeddings for sparse features and
  data,'' \emph{arXiv preprint arXiv:1901.02103}, 2019.

\bibitem{FBDLRM}
\BIBentryALTinterwordspacing
M.~Naumov and D.~Mudigere, ``Deep learning recommendation model for
  personalization and recommendation systems,'' 2019. [Online]. Available:
  \url{https://ai.facebook.com/blog/dlrm-an-advanced-open-source-deep-learning-recommendation-model/}
\BIBentrySTDinterwordspacing

\bibitem{DLRM}
\BIBentryALTinterwordspacing
M.~Naumov, D.~Mudigere, H.-J.~M. Shi, J.~Huang, N.~Sundaraman, J.~Park,
  X.~Wang, U.~Gupta, C.-J. Wu, A.~G. Azzolini, D.~Dzhulgakov, A.~Mallevich,
  I.~Cherniavskii, Y.~Lu, R.~Krishnamoorthi, A.~Yu, V.~Kondratenko, X.~Chen,
  V.~Rao, B.~Jia, L.~Xiong, and M.~Smelyanskiy, ``Deep learning recommendation
  model for personalization and recommendation systems,'' \emph{arXiv preprint
  arXiv:1906.00091}, 2019. [Online]. Available:
  \url{https://arxiv.org/abs/1906.00091}
\BIBentrySTDinterwordspacing

\bibitem{park2018deep}
J.~Park, M.~Naumov, P.~Basu, S.~Deng, A.~Kalaiah, D.~Khudia, J.~Law, P.~Malani,
  A.~Malevich, S.~Nadathur \emph{et~al.}, ``Deep learning inference in
  {Facebook} data centers: Characterization, performance optimizations and
  hardware implications,'' \emph{arXiv preprint arXiv:1811.09886}, 2018.

\bibitem{minerva}
B.~Reagen, P.~Whatmough, R.~Adolf, S.~Rama, H.~Lee, S.~K. Lee, J.~M.
  Hern{\'a}ndez-Lobato, G.-Y. Wei, and D.~Brooks, ``{Minerva: E}nabling
  low-power, highly-accurate deep neural network accelerators,'' in
  \emph{Proceedings of the ACM/IEEE 43rd Annual International Symposium on
  Computer Architecture (ISCA)}.\hskip 1em plus 0.5em minus 0.4em\relax IEEE,
  2016, pp. 267--278.

\bibitem{yolo}
J.~Redmon, S.~Divvala, R.~Girshick, and A.~Farhadi, ``You only look once:
  Unified, real-time object detection,'' in \emph{Proceedings of the IEEE
  conference on computer vision and pattern recognition}, 2016, pp. 779--788.

\bibitem{tullsen1996exploiting}
D.~M. Tullsen, S.~J. Eggers, J.~S. Emer, H.~M. Levy, J.~L. Lo, and R.~L. Stamm,
  ``Exploiting choice: Instruction fetch and issue on an implementable
  simultaneous multithreading processor,'' \emph{ACM SIGARCH Computer
  Architecture News}, vol.~24, no.~2, pp. 191--202, 1996.

\bibitem{tullsen1995simultaneous}
D.~M. Tullsen, S.~J. Eggers, and H.~M. Levy, ``Simultaneous multithreading:
  Maximizing on-chip parallelism,'' in \emph{Proceedings of the ACM SIGARCH
  Computer Architecture News}, vol.~23, no.~2.\hskip 1em plus 0.5em minus
  0.4em\relax ACM, 1995, pp. 392--403.

\bibitem{mckinsey}
\BIBentryALTinterwordspacing
C.~Underwood, ``Use cases of recommendation systems in business – current
  applications and methods,'' 2019. [Online]. Available:
  \url{https://emerj.com/ai-sector-overviews/use-cases-recommendation-systems/}
\BIBentrySTDinterwordspacing

\bibitem{wang2017deep}
R.~Wang, B.~Fu, G.~Fu, and M.~Wang, ``Deep \& cross network for ad click
  predictions,'' in \emph{Proceedings of the ADKDD'17}.\hskip 1em plus 0.5em
  minus 0.4em\relax ACM, 2017, p.~12.

\bibitem{wu2019fb_inference}
C.-J. {Wu}, D.~{Brooks}, K.~{Chen}, D.~{Chen}, S.~{Choudhury}, M.~{Dukhan},
  K.~{Hazelwood}, E.~{Isaac}, Y.~{Jia}, B.~{Jia}, T.~{Leyvand}, H.~{Lu},
  Y.~{Lu}, L.~{Qiao}, B.~{Reagen}, J.~{Spisak}, F.~{Sun}, A.~{Tulloch},
  P.~{Vajda}, X.~{Wang}, Y.~{Wang}, B.~{Wasti}, Y.~{Wu}, R.~{Xian}, S.~{Yoo},
  and P.~{Zhang}, ``Machine learning at {Facebook}: Understanding inference at
  the edge,'' in \emph{Proceedings of the IEEE International Symposium on High
  Performance Computer Architecture (HPCA)}, Feb 2019, pp. 331--344.

\bibitem{Wu:MICRO11}
C.-J. Wu, A.~Jaleel, M.~Martonosi, S.~C. Steely, Jr., and J.~Emer, ``{PACMan:
  P}refetch-aware cache management for high performance caching,'' in
  \emph{Proceedings of the 44th Annual IEEE/ACM International Symposium on
  Microarchitecture}, ser. MICRO-44, 2011.

\bibitem{Wu:ISPASS11}
C.-J. Wu and M.~Martonosi, ``Characterization and dynamic mitigation of
  intra-application cache interference,'' in \emph{Proceedings of the IEEE
  International Symposium on Performance Analysis of Systems and Software},
  ser. ISPASS '11, 2011.

\bibitem{microsoftPersonalizedRec}
\BIBentryALTinterwordspacing
X.~Xie, J.~Lian, Z.~Liu, X.~Wang, F.~Wu, H.~Wang, and Z.~Chen, ``Personalized
  recommendation systems: Five hot research topics you must know,'' 2018.
  [Online]. Available:
  \url{https://www.microsoft.com/en-us/research/lab/microsoft-research-asia/articles/personalized-recommendation-systems/}
\BIBentrySTDinterwordspacing

\bibitem{open-rec}
L.~Yang, E.~Bagdasaryan, J.~Gruenstein, C.-K. Hsieh, and D.~Estrin, ``{OpenRec:
  A} modular framework for extensible and adaptable recommendation
  algorithms,'' in \emph{Proceedings of the 11th ACM International Conference
  on Web Search and Data Mining}.\hskip 1em plus 0.5em minus 0.4em\relax ACM,
  2018, pp. 664--672.

\bibitem{you2018imagenet}
Y.~You, Z.~Zhang, C.-J. Hsieh, J.~Demmel, and K.~Keutzer, ``Imagenet training
  in minutes,'' in \emph{Proceedings of the 47th International Conference on
  Parallel Processing}.\hskip 1em plus 0.5em minus 0.4em\relax ACM, 2018, p.~1.

\bibitem{cambricon-x}
S.~{Zhang}, Z.~{Du}, L.~{Zhang}, H.~{Lan}, S.~{Liu}, L.~{Li}, Q.~{Guo},
  T.~{Chen}, and Y.~{Chen}, ``Cambricon-x: An accelerator for sparse neural
  networks,'' in \emph{Proceedings of the 49th Annual IEEE/ACM International
  Symposium on Microarchitecture (MICRO)}, Oct 2016, pp. 1--12.

\bibitem{zhu2018benchmarking}
H.~Zhu, M.~Akrout, B.~Zheng, A.~Pelegris, A.~Jayarajan, A.~Phanishayee,
  B.~Schroeder, and G.~Pekhimenko, ``Benchmarking and analyzing deep neural
  network training,'' in \emph{Proceedings of the IEEE International Symposium
  on Workload Characterization (IISWC)}.\hskip 1em plus 0.5em minus 0.4em\relax
  IEEE, 2018, pp. 88--100.

\end{thebibliography}

\end{document}